\documentclass[fleqn,11pt]{wlscirep}

\usepackage{hyperref}
\usepackage{orcidlink}
\usepackage{xcolor}
\usepackage{tikz}
\usepackage{mathtools}
\usepackage{placeins}
\usepackage{graphicx}
\usepackage{import}
\usepackage{multirow}
\usepackage[justification=justified]{caption}
\usepackage{cleveref}
\begin{document}

\title{Network Inequality through Preferential Attachment, Triadic Closure, and Homophily}
% AUTHORS
\author[1,2,3,$*$]{Jan Bachmann\orcidlink{0000-0002-6153-4714}}
\author[3,4]{Samuel Martin-Gutierrez\orcidlink{0000-0002-5685-7834}}
\author[1,2]{Lisette Esp\'in-Noboa\orcidlink{0000-0002-3945-2966}}
\author[5,6]{Nicola Cinardi\orcidlink{0000-0001-9912-713X}}
\author[3,1,$*$]{Fariba Karimi\orcidlink{0000-0002-0037-2475}}

\affil[1]{Complexity Science Hub, 1080 Vienna, Austria}
\affil[2]{Department of Network and Data Science, Central European University, 1100 Vienna, Austria}
\affil[3]{Graz University of Technology, 8010 Graz, Austria}
\affil[4]{Grupo de Sistemas Complejos, ETS de Arquitectura de Madrid, Universidad Polit\'ecnica de Madrid, 28040 Madrid, Spain}
% \affil[4]{Grupo de Sistemas Complejos, ETS de Arquitectura de Madrid, Universidad Polit\'ecnica de Madrid, Av. de Juan de Herrera, 4, 28040 Madrid}
\affil[5]{Faculty of Engineering, Free University of Bozen-Bolzano, 39100 Bolzano, Italy}
\affil[6]{Center for Computational and Stochastic Mathematics, Instituto Superior Tecnico, 1049-001 Lisbon, Portugal}
\affil[$*$]{\small{\textit{Corresponding authors email: bachmann@csh.ac.at, karimi@tugraz.at}}}

\newcommand{\jan}[1]{\textcolor{red}{Jan: #1}}

\newcommand{\lesp}[1]{\textcolor{blue}{LEEN: #1}}

\newcommand{\compound}[1]{\ensuremath{\mathrm{#1}}} % single compound LFMs like "H" or "PAH"
\newcommand{\model}[2]{\ensuremath{\mathrm{(#1,#2)}}} % model combinations like "(PAH, U)"
\newcommand{\ssim}{\ensuremath{\mathrm{sim}}} % structural similarity index
\newcommand{\semp}{\ensuremath{\mathrm{emp}}} % structural similarity index
\newcommand{\fmin}{\ensuremath{f_{\min}}}
\newcommand{\fmaj}{\ensuremath{f_{\text{maj}}}}

% Creates a section that can be referenced by its name
\makeatletter
\newcommand{\starsection}[2]{%
\section*{#1}%
\gdef\@currentlabelname{#1}% set the label name globally
\label{#2}%
\addcontentsline{toc}{section}{#1}%
}
\newcommand{\starsubsection}[2]{
  \subsection*{#1}%
  \gdef\@currentlabelname{#1}% set the label name globally
  \label{#2}%
}
\makeatother

% SI Figure and Text Mapping
\newcommand{\SIFInequity}{S1}
\newcommand{\SIFConfusion}{S2}

% Colors
\definecolor{BlueStrong}{HTML}{045993}
\definecolor{BlueWeak}{HTML}{8da5c6}
\definecolor{RedStrong}{HTML}{b30c0d}
\definecolor{RedWeak}{HTML}{db7976}

% Markers
\newcommand{\uptriangle}[1]{\tikz{\fill[#1] (0,0) -- (0.2,0) -- (0.1,0.2) -- cycle;}}
\newcommand{\downtriangle}[1]{\tikz{\fill[#1] (0,0.2) -- (0.2,0.2) -- (0.1,0) -- cycle;}}

\begin{abstract}
    \bfseries
    Inequalities in social networks arise from linking mechanisms, such as preferential attachment (connecting to popular nodes), homophily (connecting to similar others), and triadic closure (connecting through mutual contacts).
    While preferential attachment mainly drives degree inequality and homophily drives segregation, their three-way interaction remains understudied.
    This gap limits our understanding of how network inequalities emerge.
    Here, we introduce PATCH, a network growth model combining the three mechanisms to understand how they create disparities among two groups in synthetic networks.
    Extensive simulations confirm that homophily and preferential attachment increase segregation and degree inequalities, while triadic closure has countervailing effects: conditional on the other mechanisms, it amplifies population-wide degree inequality while reducing segregation and between-group degree disparities.
    We demonstrate PATCH's explanatory potential on fifty years of Physics and Computer Science collaboration and citation networks exhibiting persistent gender disparities.
    PATCH accounts for these gender disparities with the joint presence of preferential attachment, moderate gender homophily, and varying levels of triadic closure.
    By connecting mechanisms to observed inequalities, PATCH shows how their interplay sustains group disparities and provides a framework for designing interventions that promote more equitable social networks.
\end{abstract}

\maketitle

% Structure based on Communications Physics guidelines
% https://www.nature.com/documents/commsj-phys-style-formatting-guide-accept.pdf
\starsection{Introduction}{sec:introduction}
The distribution of visibility in social networks is typically highly unequal, with a small fraction of individuals accumulating a disproportionate number of links.~\cite{newman_networksintroduction_2010,mislove.etal_measurementanalysisonline_2007}
This inequality can have far-reaching consequences, as network visibility can determine access to social capital, information, and opportunities.~\cite{burt_structuralholessocial_1995,lin_socialcapitaltheory_2002}
Network structure not only affects individuals, but can also disfavor entire socio-economic groups.
Women, who are underrepresented in scientific fields like Physics,~\cite{huang.etal_historicalcomparisongender_2020,kong.etal_influencefirstmoveradvantage_2022} face additional challenges in accessing and distributing information due to their reduced visibility in the network of scientific collaborations.~\cite{zappala.etal_genderdisparitiesdissemination_2024,vasarhelyi.etal_genderinequitiesonline_2021}

Inequalities in network structure can arise from biased linking decisions.
People's tendency to form social ties with those who already have many connections, known as \textit{preferential attachment}, is one of the mechanisms that can lead to unequal degree distributions.~\cite{barabasi.albert_emergencescalingrandom_1999}
While preferential attachment is a driver of inequality in many social systems, such as academic collaboration,~\cite{merton_mattheweffectscience_1968,newman_clusteringpreferentialattachment_2001} these networks also exhibit a high degree of \textit{triadic closure}.~\cite{bachmann.etal_cumulativeadvantagebrokerage_2024,newman_structurescientificcollaboration_2001}
As a tendency for individuals to connect to friends-of-friends, triadic closure reinforces local clustering in social networks and affects network segregation.~\cite{abebe.etal_effecttriadicclosure_2022}
A third important mechanism is \textit{homophily}, the tendency to form ties with similar others,~\cite{mcpherson.etal_birdsfeatherhomophily_2001} which can lead to visibility inequalities through segregation.~\cite{karimi.etal_homophilyinfluencesranking_2018}

These mechanisms do not exist in isolation, but rather interact in complex ways.
Instead of an explicit bias towards popularity, preferential attachment can also be an implicit consequence of triadic closure, since well-connected individuals are more likely to appear as friends-of-friends.~\cite{klimek.thurner_triadicclosuredynamics_2013}
In combination with preferential attachment, homophily can push one group to the periphery of the network,~\cite{urena-carrion.etal_assortativepreferentialattachment_2023} create false perceptions of group sizes~\cite{lee.etal_homophilyminoritygroupsize_2019} or disadvantage a minority in terms of their connectivity.~\cite{karimi.etal_homophilyinfluencesranking_2018,espin-noboa.etal_inequalityinequitynetworkbased_2022}
In its interplay with triadic closure, homophily may further amplify or moderate the segregating effect of triadic closure depending on whether or not the selection of friends-of-friends is biased by homophily too.~\cite{abebe.etal_effecttriadicclosure_2022,asikainen.etal_cumulativeeffectstriadic_2020}
This could further isolate a minority group and thus restrict their access to valuable information~\cite{zappala.etal_genderdisparitiesdissemination_2024,laber.etal_effectshigherorderinteractions_2025} or exacerbate health disparities.~\cite{sajjadi.etal_structuralinequalitiesexacerbate_2025}
The interplay between these mechanisms is non-trivial, as the effect of one mechanism can be contingent on the presence of another.~\cite{abebe.etal_effecttriadicclosure_2022}

While mechanistic network models have been used extensively to study the interplay between these mechanisms~\cite{barabasi.albert_emergencescalingrandom_1999,klimek.thurner_triadicclosuredynamics_2013,holme.kim_growingscalefreenetworks_2002,abebe.etal_effecttriadicclosure_2022,asikainen.etal_cumulativeeffectstriadic_2020,karimi.etal_homophilyinfluencesranking_2018,bianconi.etal_triadicclosurebasic_2014,laber.etal_effectshigherorderinteractions_2025} they have so far not considered the three-way combination of preferential attachment, triadic closure, and homophily or their impact on network inequality in the presence of a minority group.
We introduce PA--TC--H, a network growth model extending a stream of models originating from the preferential attachment (PA) Barab\'asi-Albert model~\cite{barabasi.albert_emergencescalingrandom_1999} by combining existing triadic closure (TC)~\cite{holme.kim_growingscalefreenetworks_2002,bianconi.etal_triadicclosurebasic_2014} and homophily (H)~\cite{karimi.etal_homophilyinfluencesranking_2018} extensions.
By tuning their parameters, mechanistic models can be used to study the impact of biased link formation mechanisms on network inequality,~\cite{karimi.etal_homophilyinfluencesranking_2018,espin-noboa.etal_inequalityinequitynetworkbased_2022} network interventions,~\cite{neuhauser.etal_improvingvisibilityminorities_2023} and to create diverse synthetic networks.

Our comprehensive analysis shows that homophily and preferential attachment can segregate a minority from the majority group in the network, and increase degree inequality within each group and among all nodes.
Homophily also increases network inequity, the degree disparity between the minority and majority groups.
Triadic closure mitigates segregation and inequity only when the selection among friends of friends is not biased by the other two mechanisms.
To identify the mechanisms and parameters that best explain the varying gender inequalities in scientific collaboration and citation, we apply likelihood-free inference to three empirical networks.
We find that the empirically observed inequalities are best explained by a model where both global search and local triadic closure links are biased by preferential attachment and homophily.
Our results suggest that mitigating inequalities based on link formation can only be achieved when keeping all mechanisms and other inequalities in mind, as isolating one can lead to unintended consequences.

\starsection{Results}{sec:patch}
% - modeling enables
% 	- causality
% 	- what-if analysis (interventions)
% - extension of PA + H model
% 	- $N$ total number of nodes
% 	- $m$ edges formed per newly joining node
% 	- initiation
% 		- $m$ fully connected nodes
% 			- justification: no isolates, minimum degree $m$
% 	- remaining $N-m$ nodes join one after the other and each link to $m$ previously added nodes
% 		- how the target node is chosen is biased by the respective model choice
% - addition of triadic closure as target candidate limitation
% 	- with probability $l\in[0,1]$ we form a local link
% 		- limited to friends-of-friends
% 	- otherwise, all existing nodes are valid partners
% - among chosen target set the selection is
% 	- homophily
% 		 - mixing matrix $H = [h_{mm} = h, h_{mM} = 1 - h; h_{Mm} = 1 - h, h_{MM} = h]$
% 		 - $P_{ij} = H_{m_im_j}$
% 	 - preferential attachment + homophily
% 		 - $P_{ij} = k_j  H_{m_im_j} / \sum_{n < i} k_n H_{m_im_n}$
% 		 - note that $h=0.5$ is preferential attachment without homophily
%  - simulation for $N=5\,000$ and $m=2$, higher $m$ in SI, $50$ realizations
% 	 - higher $m$ introduces multiplicity bias
% 		 - one can be friends of multiple friends
% 		 - choice to include as bias
\begin{figure}[ht]
    \centering
    \includegraphics[width=\textwidth]{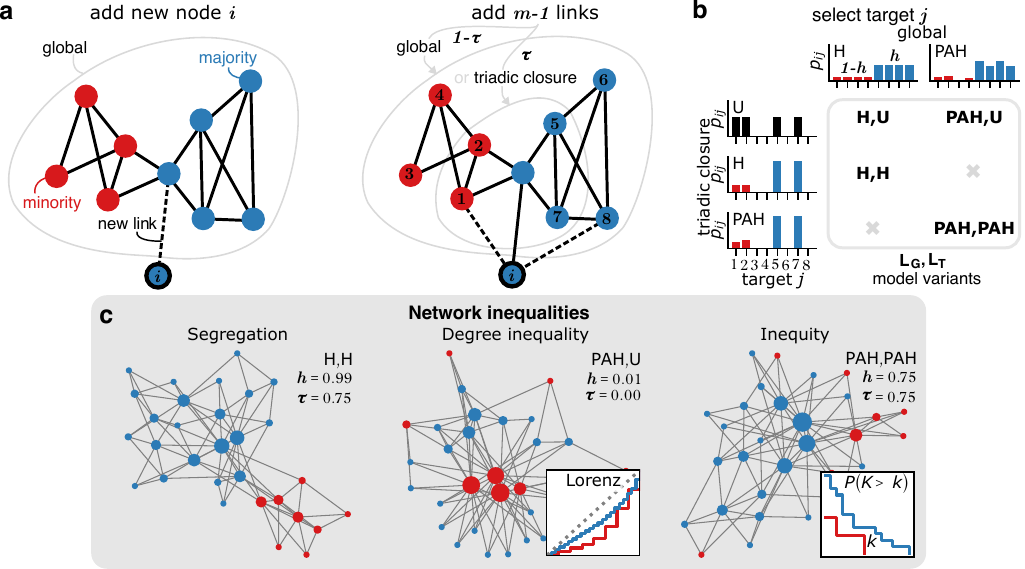}
    \caption{\label{fig:patch_model}\textbf{PATCH networks}.
    (\textbf{a}) Network growth: A new node $i$ joins the network and is assigned to the majority group with probability $1 - \fmin$, or the minority otherwise.
    For its first link, $i$ selects a target node globally, from all existing nodes (gray outline).
    Subsequently, $i$ creates $m-1$ additional links.
    For each new link, $i$ connects to a target either globally with probability $1-\tau$, as in the first link, or through triadic closure, limiting target nodes to friends of friends $j \in \{1, 2, 5, 7\}$ (inner gray outline).
    (\textbf{b}) Linking mechanisms: Among the available triadic closure target nodes $j$, $i$ chooses based on one of three mechanisms $p_{ij}$: Targets are chosen uniformly at random without any bias (\compound{U}), solely based on group membership with homophily (\compound{H}), or by combination of degree-based preferential attachment and homophily (\compound{PAH}).
    While the mechanisms may vary between global and triadic closure links, we consider only model variants \model{L_G}{L_T} in which triadic closure is unbiased ($ \compound{L_T} \coloneqq \compound{U}$) or biased in the same way as the global selection $\compound{L_T} \coloneqq \compound{L_G} \in \{ \compound{H}, \compound{PAH} \}$.
    Homophily is controlled by a parameter $h$ favoring in-group links for $h>0.5$ or out-group links for $h<0.5$.
    (\textbf{c}) Network inequalities: Three PATCH networks with high segregation, unequal degrees, and with one group dominating the other, illustrating the three network outcomes we analyze.}
\end{figure}

PATCH is based on the growth mechanism of the Barab\'asi-Albert network model which sequentially adds a total of $N$ nodes to the network, each linking to $m$ previously added nodes (\Cref{fig:patch_model}).~\cite{barabasi.albert_emergencescalingrandom_1999}
Following existing triadic closure variants of this model, a probability $\tau$ determines whether the pool of potential target nodes for each new link is restricted to friends-of-friends or includes all nodes in the existing network.~\cite{bianconi.etal_triadicclosurebasic_2014,holme.kim_growingscalefreenetworks_2002}
We refer to the former as triadic closure links and the latter as global links.
To choose among the target nodes, we consider three different mechanisms: unbiased selection (\compound{U}), homophily (\compound{H}), and preferential attachment with homophily (\compound{PAH}).
Following another common extension, we control homophily by a parameter $h \in (0,1)$ with a preference for in-group links for $h > 0.5$ and out-group connections for $h < 0.5$.~\cite{karimi.etal_homophilyinfluencesranking_2018}
To model social group dynamics, we assume each node belongs to either a minority or majority group, with the minority group representing a fraction $\fmin < 0.5$ of the $N$ nodes.
In the presence of preferential attachment (\compound{PAH}), the probability of forming a link is proportional to the degree of the target node, matching the baseline model.~\cite{barabasi.albert_emergencescalingrandom_1999}
The formal definitions and parameterizations of all mechanisms are detailed in~\nameref{sec:methods}.
Since prior literature suggests a reduced importance of homophily when choosing triadic closure targets,~\cite{kossinets.watts_originshomophilyevolving_2009} we vary mechanisms between global and triadic closure links.
To this end, we distinguish four global ($\mathrm{L_G}$) and triadic closure ($\mathrm{L_T}$) link mechanism combinations $\model{\mathrm{L_G}}{\mathrm{L_T}} \in \{\model{H}{U},\model{H}{H},\model{PAH}{U},\model{PAH}{PAH}\}$.
We then simulate the model with $N=5\,000$ nodes and $m=3$ initial links, and a minority fraction of $\fmin = 0.2$ while varying $h$ and $\tau$ as control parameters.

\subsection*{\label{sec:segregation} Network segregation}
\begin{figure}[ht]
    \centering
    \includegraphics[width=\textwidth]{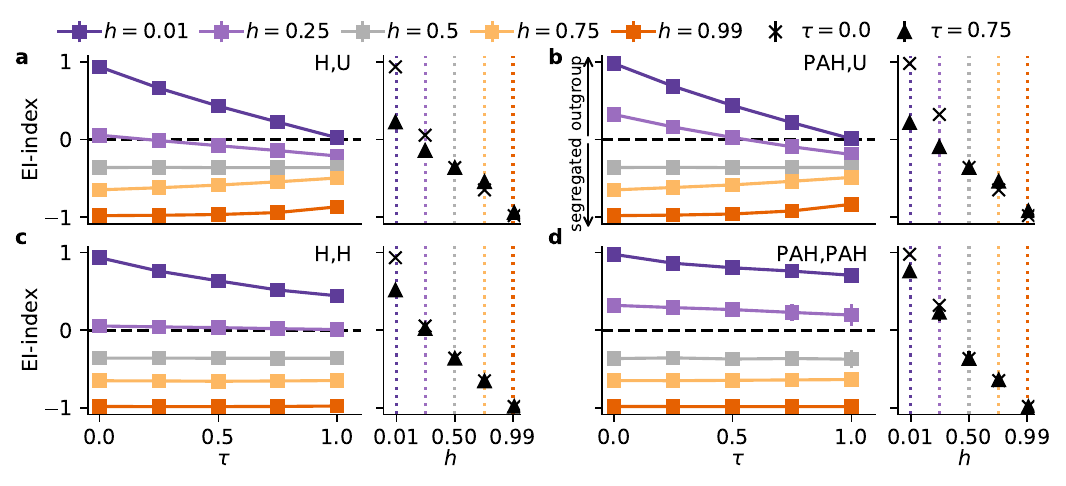}
    \caption{\label{fig:segregation}\textbf{Homophily drives segregation}.
    We vary the triadic closure probability $\tau$ and homophily $h$ (x-axes of the wider and shorter subplots, respectively) and the link formation combinations (\textbf{a}--\textbf{d}) to measure the network segregation by the EI-index (y-axis).
    Negative values indicate segregation, neutral values indicate mixing, and positive values indicate outgroup linking.
    (\textbf{a}) Due to the group size imbalance, the network is slightly segregated ($EI < 0$) even in the neutral case ($h=0.5$ and $\tau=0.0$).
    (\textbf{a}--\textbf{d}) Segregation is mainly driven by homophily $h$ (from purple to orange).
    Under homophilic preferences ($h > 0.5$) the network becomes segregated, while heterophilic tendencies ($h < 0.5$) lead to a high inter-group connectivity.
    The effect of triadic closure is not straightforward.
    (\textbf{a} and \textbf{b}) If the selection is unbiased ($\mathrm{L_T} = \compound{U}$), it moderates strong segregation or strong mixing towards the neutral state (gray color).
    (\textbf{c} and \textbf{d}) If the selection is biased ($\mathrm{L_T} = \mathrm{L_G}$) and nodes prefer heterophilic linking ($h < 0.5$), triadic closure slightly shifts the network towards more in-group linking.
    Under homophily ($h > 0.5$), triadic closure has no effect.}
\end{figure}
% 1. Here start with the finidings from previous work (eg AKsikain found that TC always amplifies \compound{H})
% 2. [Homophily can result from structural factors or from choice (McPherson et al. 2001), but either may generate pressures for greater inequality given the presence of network effects.]
% 3. [Additional research is needed, as well, on the relationship between homophily and triadic closure—under what conditions, for example, individuals tend to segregate their network based on the traits they share with different alters, as opposed to bring together alters with different characteristics (possibly creating bridges) (but see Kossinets & Watts 2009).]
\textit{Network segregation} can isolate a group from accessing information, opportunities or support,~\cite{karimi.etal_homophilyinfluencesranking_2018,zappala.etal_genderdisparitiesdissemination_2024} thus driving socio-economic inequalities.~\cite{calvo-armengol.jackson_effectssocialnetworks_2004}
The observed mixing preferences between various groups in a network depend not only on the homophily preference of individual actors, but also on the local network structure that they are embedded in and the presence of other link formation mechanisms.~\cite{asikainen.etal_cumulativeeffectstriadic_2020,sajjadi.etal_unveilinghomophilypool_2024}
% We distinguish between the individual homophily preference as choice homophily (parameterized by $h$) and the outcome homophily as measured by the segregation of the final network.
The effect of triadic closure on network segregation is not straightforward.
While some studies suggest that triadic closure can moderate network segregation,~\cite{abebe.etal_effecttriadicclosure_2022} others argue that it can amplify it.~\cite{asikainen.etal_cumulativeeffectstriadic_2020}
The main distinction is the presence of homophily in the selection of triadic closure targets, that is, whether nodes prefer to link to friends of friends with similar attributes.~\cite{abebe.etal_effecttriadicclosure_2022}
Empirical studies suggest that the effect of homophily, while still present, decreases with the proximity of target nodes.~\cite{kossinets.watts_originshomophilyevolving_2009}
In combination, this suggests that triadic closure can moderate segregation.
We investigate the effect of triadic closure on network segregation with or without a biased selection of triadic closure targets by choosing $\mathrm{L_T}=\mathrm{L_G}$ or $\mathrm{L_T}=\compound{U}$, respectively.

We measure the segregation of the simulated network by the fraction of the frequency of internal links $I$ to external links $E$ as
\begin{equation}
    EI = \frac{E - I}{E + I}
\end{equation}
ranging from $-1$ for fully segregated networks to $+1$ for networks in which members of one group exclusively link to the other group (\Cref{fig:segregation}).~\cite{krackhardt.stern_informalnetworksorganizational_1988}
Varying the homophily $h\in\{0.01, 0.25, 0.50, 0.75, 0.99\}$ and the triadic closure probability $\tau \in \{0.0, 0.25, 0.5, 0.75, 1.0\}$, we identify the strongest effect with increasing $h$.
Homophily drives network segregation, regardless of any combination of link formation mechanisms.

We observe a moderating effect of triadic closure, driving the system towards a slightly segregated state which aligns with the neutral case (\Cref{fig:segregation}.\textbf{a} with $h=0.5$).
Surprisingly, triadic closure does not seem to amplify segregation, even when it is biased itself (\model{H}{H} and \model{PAH}{PAH}, bottom row).
There is no amplification of segregation in the network, as the EI-index remains constant over increasing $\tau$.
Only in the case of heterophily ($h < 0.5$, purple coloring) we observe a decrease in the EI-index, indicating that triadic closure counteracts cross-group mixing by favoring within-group ties.
This is surprising, given that triadic closure, too, is biased towards heterophily.
Together, this adds nuances to our understanding of the amplification of segregation through triadic closure: (i) in agreement with existing studies,~\cite{abebe.etal_effecttriadicclosure_2022} triadic closure moderates segregation when it is unbiased.
(ii) It does not amplify segregation under homophily ($h>0.5$) and this is not explained by the presence of preferential attachment (see \model{H}{H} in \Cref{fig:segregation}).
% (iii) In the heterophilic case ($h<0.5$), TC increases segregation, regardless of whether it is biased or not.
(iii) In the heterophilic case ($h<0.5$), it moderates the abundance of outgroup links towards a well-mixed network, regardless of whether it is biased or not.

%% Summarized results (slide from DPG presentation in Berlin - Samuel)
% -Inequality is mostly driven by Prefferential Attachment – PA exacerbates inequality when acting at the local level.
% -Triadic Closure moderates network segregation in most situations (~as in Abebe et al.).
% -Triadic Closure amplifies network segregation when homophily is the only global and local mechanism (~as in Asikainen et al.).
% -Homophily is the main driver of inequity.
% -Both TC and PA tend to reduce inter-group inequity, even when TC is biased!

%% Same results written in a paragraph with decent flow:
% As shown in Figure XX, when we include PA in the model, inequality (Gini) always increases, regardless of whether we include it at the global or at the local scale. Furthermore, if we include random local links to a network with global PA, we reduce PA. Overall, these patterns indicate that PA is the main driver of inequality. Panels XX through XX in Figure XX show that triadic closure moderates network segregation in most cases, as it always brings the EI measure closer to the neutral 0 line. However, in specific cases, such as when we have both global and local homophily linkage, triadic closure (local linking) always reduces EI, indicating an increase of outcome homophily \emph{regardless of whether connections are homophilic or heterophilic}. Finally, homophily seems to be the main driver of inequity (inter-group degree disparities), because the inclusion of any of the other tie formation mechanism (PA or TC) always reduces inequity.

\subsection*{\label{sec:inequality}Inequality in networks}
% - consider degree as proxy of social capital
% 	- _move reference about minority visibility here?_

\begin{figure}[ht]
    \centering
    \includegraphics[width=\textwidth]{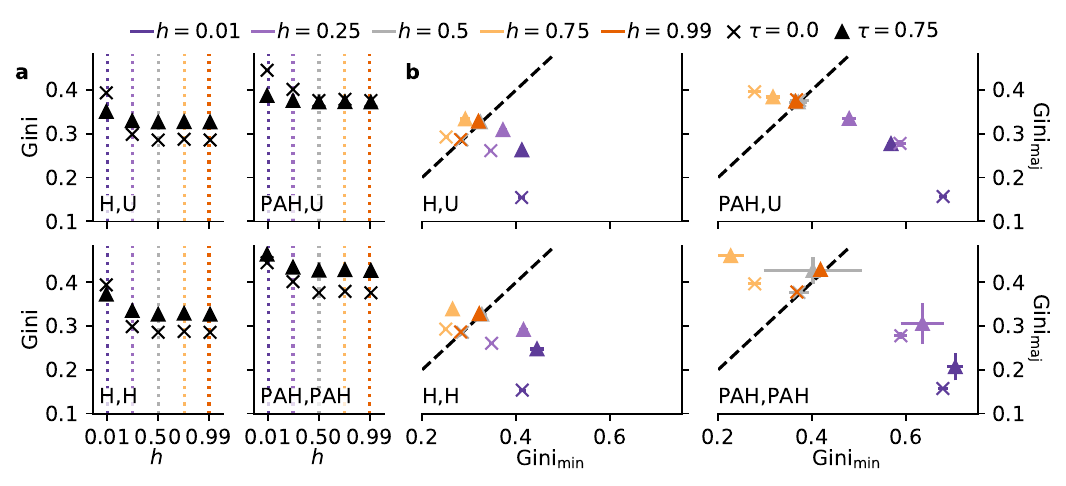}
    \caption{\label{fig:inequality}\textbf{Preferential attachment drives degree inequality}.
    (\textbf{a}) Measuring global degree inequality by the Gini coefficient, we identify the strongest effect by the presence of preferential attachment \model{PAH}{U} and \model{PAH}{PAH}.
    In heterophily ($h < 0.5$), only few nodes retain most of the available links, even in the absence of preferential attachment.
    While the effect of triadic closure depends on the other mechanisms, it exacerbates degree inequality in most cases.
    (\textbf{b}) In-group degree inequality behaves similarly.
    Degree inequality is higher for the majority group under homophily.
    While extreme homophily equalizes the difference between inequalities, it is strongest for the minority under strong heterophily.
    This is between group disparity is further exacerbated by preferential attachment.}
\end{figure}
% - measure Gini coefficient
% - plot degree distributions and overview of Gini coefficient across models
% - show that triadic closure mostly exacerbates degree inequality
% 	- potential reason: friendship paradox
% - only when local choice is unbiased (no PA and $h=0.5$) and global selection has PA it reduces Gini because fewer links go to popular nodes
Social networks tend to exhibit strong \textit{degree inequalities} as the number of links nodes accumulate is distributed unevenly.~\cite{newman_networksintroduction_2010}
The Gini coefficient is a common measure to quantify this inequality, ranging from zero for a perfectly equal distribution to one for a network in which a single node holds all connections.~\cite{espin-noboa.etal_inequalityinequitynetworkbased_2022}
Prior literature has identified preferential attachment,~\cite{barabasi.albert_emergencescalingrandom_1999} especially in combination with extreme heterophily or homophily~\cite{espin-noboa.etal_inequalityinequitynetworkbased_2022} as a key driver of degree inequality in social networks.
At the same time, it can act as sub-linear preferential attachment (i.e., the connection probability is $p_{ij} \propto k_j^\alpha$ with $\alpha < 1$),~\cite{klimek.thurner_triadicclosuredynamics_2013} potentially exacerbating degree inequality.

The Gini coefficient is mostly driven by the presence of preferential attachment in our model (\Cref{fig:inequality}.\textbf{a}).
Preferential attachment is known to create right-skewed degree distributions~\cite{barabasi.albert_emergencescalingrandom_1999} which are inherently unequal.
In the absence of preferential attachment and triadic closure (left column and $\tau=0$), the Gini coefficient generally decreases with homophily $h$, as nodes tend to link more and more to similar others.
This is most likely because in the heterophilic setting ($h < 0.5$), a few minority nodes receive most links of the majority group.~\cite{karimi.etal_homophilyinfluencesranking_2018}
Our results show that this is true even in the absence of preferential attachment; an increase in homophily would motivate minority nodes to link to other minority nodes, decreasing the skewness of the distribution.

%The effect of triadic closure on degree inequality is not straightforward.
Apart from the extreme heterophily scenario, triadic closure exacerbates degree inequality, which is in line with the sub-linear preferential attachment it creates.~\cite{klimek.thurner_triadicclosuredynamics_2013}
This is likely due to the friendship paradox, which states that high-degree nodes are overrepresented among the friends of existing neighbors.~\cite{feld_whyyourfriends_1991}
To distinguish the two types of preferential attachment, we define the explicitly modelled preference to connect to popular nodes as \textit{choice preferential attachment} and the effect of triadic closure as \textit{induced preferential attachment}.

When nodes globally choose based on choice preferential attachment, but are locally unbiased \model{PAH}{U}, or under extreme heterophilic preference ($h=0.01$) without any choice preferential attachment (\model{H}{U} and \model{H}{H}), the Gini coefficient decreases with $\tau$ (\Cref{fig:inequality}.\textbf{a}).
In the former case, triadic closure reduces the number of links drawn based on preferential attachment.
Compared to choice preferential attachment, the induced preferential attachment is sub-linear, which then reduces the overall degree inequality.
In the second scenario, triadic closure most likely balances out the degree inequality by giving more links to the larger majority group (compare reduced segregation in~\Cref{fig:segregation}.\textbf{a} and \textbf{c}) which distributes links across more individuals.%~\cite{karimi.etal_homophilyinfluencesranking_2018}

Groups may not be affected by degree inequality in the same way.
% Degree inequalities are not necessarily distributed equally across groups, but might be stronger in one of the two groups.
Measuring the inequality per group by the Gini coefficient of the degree distribution of the minority $\mathrm{Gini_{min}}$ and majority group $\mathrm{Gini_{maj}}$ separately, we observe a negative correlation between the two coefficients (\Cref{fig:inequality}.\textbf{b}).
If one group has a high degree inequality, the other group tends to have a more evenly distributed degree visibility.
This further indicates that the global degree inequality is mainly driven by the more unequal group.
Namely, the degree distribution of the majority group is more unequal under intermediate homophilic preferences ($h = 0.75$), while the minority group is more unequal with heterophilic mixing ($h < 0.5$).
At the same time, the other group shows less inequality than the neutral case.
This is especially pronounced when both global and triadic closure links are biased by preferential attachment and homophily \model{PAH}{PAH}.
While one group shows the most extreme inequality, the other one remains almost equal.
Only in the absence of homophily ($h=0.5$) or extreme homophily ($h=0.99$), we observe identical, but high Gini coefficients for both groups and all model variants.

\subsection*{\label{sec:inequity}Inequity in networks}
% - measure stochastic dominance of minority over majority
% 	- considers whole distribution, not driven by extremes
% - plot
% 	- refer to degree distribution plot
% 	- show model comparison plot with inter-group versus (population level or intra-group?)
% 	- show SD lines over homophily
% - triadic closure neutralizes inequality towards neutral value
% 	- except for global unbiased, local bias
\begin{figure}[t]
    \centering
    \includegraphics[width=\textwidth]{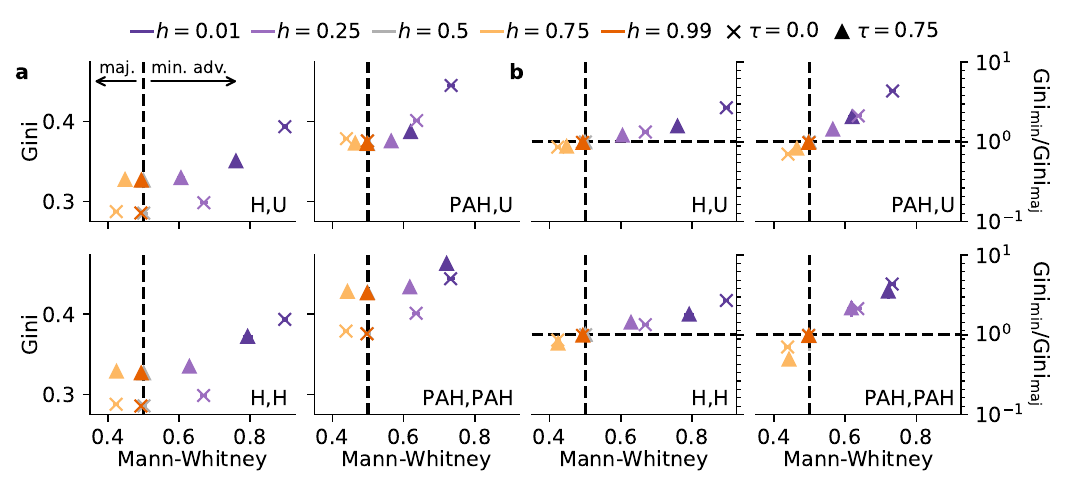}
    \caption{\label{fig:inequity_inequality}\textbf{Inequity and its relationship with inequality}. (\textbf{a}) We measure inequity as the capability of one group to accumulate more links than the other by the Mann-Whitney test statistic.
    Values below $0.5$ indicate that the majority group are advantaged in degree visibility, while values above $0.5$ indicate the opposite.
    Homophily drives inequity while unbiased triadic closure and the presence of preferential attachment moderate it.
    We reproduce in our undirected network model the U-shaped relation between inequality and inequity originally found in a similar, directed model.~\cite{espin-noboa.etal_inequalityinequitynetworkbased_2022}
    One group being more visible than the other always comes with a higher degree inequality among all nodes.
    (\textbf{b}) This inequality is always higher for the advantaged group.
    In the heterophilic case ($h < 0.5$), the minority group is disadvantaged and shows a higher inequality.
    Under intermediate homophily levels ($h = 0.75$), the advantage of the majority group matches their increased degree inequality.
    This indicates that an advantage in degree visibility favors only a few nodes from one group.
    Only unbiased triadic closure consistently neutralizes this effect.
    }
\end{figure}

Our findings of asymmetric inequalities are linked to \textit{inequity}, the advantaged position of one group over the other. %~\cite{espin-noboa.etal_inequalityinequitynetworkbased_2022}
We measure the inequity between the two groups as the probability that a randomly chosen node from the minority group has a higher degree than a randomly chosen node from the majority group.
This measure reflects the common language interpretation~\cite{mcgraw.wong_commonlanguageeffect_1992} of the Mann-Whitney test statistic.~\cite{mann.whitney_testwhetherone_1947}
Values above $50\%$ indicate that the minority group has a higher degree visibility than the majority group, while values below $50\%$ indicate the opposite.

In agreement with previous literature,~\cite{karimi.etal_homophilyinfluencesranking_2018,espin-noboa.etal_inequalityinequitynetworkbased_2022} we identify homophily $h$ as the main driver of inequity (changes in the x-axis of~\Cref{fig:inequity_inequality}).
In the intermediate homophily case ($h = 0.75$), the majority group achieves greater visibility than the minority group.
The opposite is true under heterophily ($h < 0.5$), when the minority group dominates the majority.
Triadic closure and preferential attachment have a moderating effect on inequity instead, as they reduce the inequity towards the neutral value of $50\%$.

The relationship between inequity and degree inequality is U-shaped, as previously observed in a similar, but directed network model.~\cite{espin-noboa.etal_inequalityinequitynetworkbased_2022}
Whenever one group is more visible than the other, the degree inequality among all nodes is higher.
This occurs irrespective of the presence of preferential attachment or triadic closure and it is strongest in the heterophilic case ($h < 0.5$), where the minority group is advantaged.
In~\Cref{fig:inequity_inequality}.\textbf{b}, we link inequity to the intra-group inequality summarized by the ratio of the minority and majority groups' Gini coefficients $\mathrm{Gini_{min}/Gini_{maj}}$.
We observe a positive linear trend between inequity and the intra-group inequality in all model variants.
The advantaged group always has a higher degree inequality.
The presence of preferential attachment shifts this effect away towards lower inequity but higher degree inequality among the advantaged group.
The advantage in degree visibility favors only a few nodes from the dominant group while most nodes remain less visible.

\begin{table}[t]
    \centering
    \small
    
\renewcommand{\arraystretch}{1.2}
\begin{tabular}{lllccccc}
\toprule
&&&
\multicolumn{1}{c}{\textbf{Segregation} (EI)} &
\multicolumn{3}{c}{\textbf{Degree inequality} (Gini)} &
\multicolumn{1}{c}{\textbf{Inequity} (MW)} \\
% \cmidrule(lr){2-2}\cmidrule(lr){3-5}\cmidrule(lr){6-6}
Mechanism &
\multicolumn{2}{l}{Parameterization} &
&
All &
Min &
Maj &
\\
\midrule

{\bf Preferential attachment} &
% \multicolumn{1}{l}{$\rm \boldsymbol{*} \xrightarrow{L_{G,T}} PAH$} &
set $\rm L_{G,T}$ to \compound{PAH}&
&
% \multicolumn{2}{c}{--} &
--
 &
\uptriangle{RedStrong} &
\uptriangle{RedStrong} &
\hspace{11pt} ${\uptriangle{RedWeak}^{\rm hom}}$ &
% \multicolumn{2}{c}{--}
--
\\
\addlinespace

\multirow{1}{*}{\bf Triadic closure} &
\multicolumn{2}{l}{unbiased, increase $\tau$} &
% $0 \xrightarrow{\tau} 1$ &
% unbiased &
$\uptriangle{black}\!\downtriangle{black}$ &
\hspace{6pt} $\downtriangle{BlueWeak}_{\rm PA}$ &
% $\uptriangle{RedWeak}\!\downtriangle{BlueWeak}_{\rm PA}$ &
% -- &
% -- &
-- &
-- &
$\uptriangle{black}\!\downtriangle{black}$ \\

&
% $0 \xrightarrow{\tau} 1$ &
\multicolumn{2}{l}{biased, increase $\tau$}&
\hspace{6pt} $\uptriangle{gray}\!\downtriangle{gray}_{\rm het}$ &
% -- &
% -- &
% \uptriangle{RedWeak}
-- &
-- &
$\uptriangle{RedWeak}$ &
--
\\
\addlinespace

\multirow{1}{*}{\bf Homophily} &
% $0.50 \xrightarrow{h} 0.01$ &
\multicolumn{2}{l}{towards heterophily}&
% heterophily &
\uptriangle{RedStrong} &
$\uptriangle{RedWeak}$ &
$\uptriangle{RedWeak}$ &
$\downtriangle{BlueWeak}$ &
\uptriangle{RedStrong}
\\

&
% $0.50 \xrightarrow{h} 0.75$ &
\multicolumn{2}{l}{moderate homophily} &
\downtriangle{BlueStrong} &
-- &
$\downtriangle{BlueWeak}$ &
$\uptriangle{RedWeak}$ &
\downtriangle{BlueStrong} \\
 &
 % $0.75 \xrightarrow{h} 0.99$ &
 \multicolumn{2}{l}{extreme homophily} &

\downtriangle{BlueStrong} &
$\uptriangle{RedWeak}$ &
$\downtriangle{BlueWeak}$ &
$\downtriangle{BlueWeak}$ &
$\uptriangle{gray}\!\downtriangle{gray}$ \\
\bottomrule
\end{tabular}
\vspace{0.5pt}
\begin{minipage}{0.97\linewidth}\small
    -- No or mixed effect\;
    \uptriangle{RedWeak} Metric increase\;
    \downtriangle{BlueWeak} Decrease\;
    $\uptriangle{gray}\!\downtriangle{gray}$ Neutralization\;
    $\uptriangle{RedStrong}\!\downtriangle{BlueStrong} / \uptriangle{black}\!\downtriangle{black}$ Main up- or downwards driver/neutralizer
\end{minipage}
% The effect of preferential attachment (PA) summarizes model variant comparisons by replacing \compound{H} or \compound{U} by \compound{PAH}: $\model{L_G=H}{L_T=U} \rightarrow \model{PAH}{U}$, $\model{H}{H} \rightarrow \model{PAH}{PAH}$, and $\model{PAH}{U} \rightarrow \model{PAH}{PAH}$.
% The contribution of triadic closure, either unbiased ($\rm L_T = U$) or biased ($\rm L_T = L_G$), is considered by increasing $\tau$ on $0 \xrightarrow{\tau} 1$.
% For homophily, we distinguish between changes from neutrality towards heterophily ($0.50 \xrightarrow{h} 0.01$) or moderate homophily ($0.50 \xrightarrow{h} 0.75$), and from moderate to extreme homophily ($0.75 \xrightarrow{h} 0.99$).
% Single triangles indicate high or low metric values and double triangles symbolizes neutralizing changes towards equality.
% ffects may be general, or conditional on the presence of preferential attachment (`PA'), heterophily (`het'), or homophily (`hom').
% We mark the strongest driver or mitigator per network inequality in red and green, respectively.}
    \caption{
        \label{tab:summary}
        \textbf{Qualitative summary of outcomes}.
        Linking mechanisms (PA, TC, H) %(rows)
        affect network segregation (EI), degree inequality (Gini) and inequity (MW). % (columns).
        Single triangles and color indicate high or low metric values and double triangles symbolize neutralizing changes towards equality.
        Effects may be general, or depend on preferential attachment (`PA'), heterophily (`het'), or homophily (`hom').
        Color shades indicate the effect strength, with the darkest shade marking the mechanism that contributes the strongest to the respective inequality.
        PA does not affect segregation or inequity, but it skews the degree distribution of minorities, and of majorities only under homophily.
        The effects of triadic closure depend on whether it is biased (node choice follows the global mechanism, $ \compound{L_T}=\compound{L_G} $) or unbiased (uniform choice, $\compound{L_T} = \compound{U}$).
        In the unbiased case, triadic closure reduces segregation and inequity.
        In the biased case, it neutralizes segregation only in heterophilic networks.
        Homophily segregates networks under homophilic preferences and towards outgroup linking under heterophily.
        Moving from its absence to heterophily or to moderate homophily creates inequity, favoring the minority or majority group, respectively.
        The favored group experiences higher degree inequality, indicating that only few nodes benefit from the advantage.}
\end{table}

PATCH produces a wide variety of networks and inequalities by variation of its mechanisms.
\Cref{tab:summary} summarizes the main effects and interplays of the three mechanisms on segregation, inequality, and inequity.
Given an observed network, the table maps its inequalities to contributing mechanisms, providing explanations and guiding the implementation of mitigation strategies.

\subsection*{\label{sec:empirical_networks}Network inequality in academic networks}
\FloatBarrier

\begin{figure}[t]
    \centering
    \includegraphics[width=\textwidth]{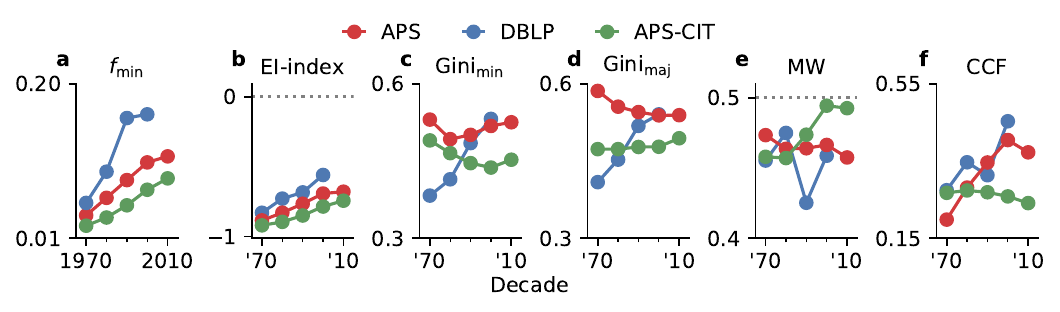}
    \caption{\label{fig:empirical}\textbf{Gender inequalities in scientific networks} over five decades. Each subplot shows an aggregate network statistic summarizing observed network inequalities and informing the PATCH model parameter inference.
    (\textbf{a}) While the fraction of women as the minority group $\fmin$ is increasing consistently over time for all datasets, it remains well below parity.
    (\textbf{b}) Segregation as measured by the EI-index increases simultaneously.
    (\textbf{c}--\textbf{d}) Within-group inequality and how it is distributed among men and women varies by dataset and time.
    APS collaboration exhibits the strongest degree inequality in the 1970s, decreasing consistently only among men.
    Degree inequality in APS citations shows a reversed trend instead.
    Collaboration among DBLP authors is quickly becoming more unequal in the entire population.
    (\textbf{e}) Despite their growing representation women are disadvantaged in all datasets and decades except in citations starting in the 1990s, as measured by the Mann-Whitney statistic (MW).
    (\textbf{f}) The average clustering coefficient (CCF), measuring the fraction of connected neighbors out of all possible pairs of neighbors, increases in collaboration and decreases in citation.}
\end{figure}

We apply PATCH to three empirical networks, considering scientific collaborations within journals of the American Physical Society (APS),~\cite{americanphysicalsociety_americanphysicalsociety_2023} citations among APS publications (APS-CIT)~\cite{americanphysicalsociety_americanphysicalsociety_2023} and scientific collaborations within the Computer Science Bibliography (DBLP).~\cite{dblpteam_dblpcomputerscience_2016}
These datasets reflect the network of collaborations and citation practices among researchers in the fields of Physics and Computer Science, respectively; fields in which women are historically underrepresented.~\cite{huang.etal_historicalcomparisongender_2020,barnes.etal_edgeinterventionscan_2025}
Collaboration networks are moderately segregated by gender~\cite{sajjadi.etal_unveilinghomophilypool_2024,karimi.oliveira_inadequacynominalassortativity_2023,jadidi.etal_genderdisparitiesscience_2017} and highly clustered,~\cite{newman_structurescientificcollaboration_2001} suggesting the presence of triadic closure and homophily in the decision-making process of forming collaborations (see~\nameref{sec:methods} for details on how the networks are created).
The number of collaborators per author is highly unequal, with a few authors having many collaborators and most authors having only a few collaborators~\cite{newman_structurescientificcollaboration_2001} which can be attributed both to preferential attachment and triadic closure.~\cite{newman_clusteringpreferentialattachment_2001}
Identifying the presence of biases, such as preferential attachment, gender homophily or triadic closure in the formation of collaborations is crucial to reduce the inherent inequalities in science in general~\cite{merton_mattheweffectscience_1968} and between men and women specifically.~\cite{huang.etal_historicalcomparisongender_2020,lariviere.etal_bibliometricsglobalgender_2013}
With whom an author gets to collaborate can greatly influence their visibility and career prospects.~\cite{li.etal_untanglingnetworkeffects_2022,sarigol.etal_predictingscientificsuccess_2014,sekara.etal_chaperoneeffectscientific_2018,dies.etal_forecastingfacultyplacement_2025}
As accumulated citations are often used to measure scientists' impact, gender homophilic citation practices~\cite{dworkin.etal_extentdriversgender_2020} could drive the inequalities in citations received.~\cite{huang.etal_historicalcomparisongender_2020,jaramillo.etal_systematiccomparisongender_2025}
Although PATCH is generally agnostic to the context it is applied to, incorporating preferential attachment, triadic closure, and (gender) homophily makes it a valuable tool to understand the role of latent behavioral biases on the emergence of inequalities.

Considering a roughly equal representation of men and women in the underlying general population, the fraction of women in Physics and Computer Science increased only slowly between 1970 and 2010 (\Cref{fig:empirical}.\textbf{a}).
Even DBLP, which shows the highest increase, remains well below 20\% of women in the 2000s.
Women's reduced representation in the citation network is likely linked to their lower productivity which is due to higher dropout rates, shorter careers,~\cite{huang.etal_historicalcomparisongender_2020} and unequal parental obligations.~\cite{morgan.etal_unequalimpactparenthood_2021}
As women's representation increases, we observe an upwards trend in across-gender collaborations and citations over time as indicated by an increasing EI index (\Cref{fig:empirical}.\textbf{b}).
Still, the gender segregation is evident by the EI-index starting almost at its minimum of -1.
As discussed in~\Cref{fig:segregation}, this is not necessarily indicative of underlying behavioral differences, i.e., homophily, but can partly be due to women's smaller group size.

While men and women share a strong degree inequality, the three datasets show distinctive trend variations (\Cref{fig:empirical}.\textbf{c} \& \textbf{d}).
In the citation network, inequality levels appear to decrease for women, but increase for men.
In contrast, APS shows the reversed effect and degree inequalities in DBLP sharply increase for both groups.
Considering the across-gender inequity, we observe persistently lower degrees for women in terms of number of co-authors and citations in line with prior research (\Cref{fig:empirical}.\textbf{e}).~\cite{huang.etal_historicalcomparisongender_2020}
Only the citation network develops towards the neutral state in which men and women share the same degree visibility.
Note however that all measures are computed irrespective of the link direction and thus incorporate both outgoing and incoming citations.

Besides the presented network inequality statistics, we also consider the average local clustering coefficient (CCF) of the network to better capture the effect of triadic closure.
The observed CCF is generally higher for the collaboration networks following the expected prevalence of triadic closure,~\cite{newman_structurescientificcollaboration_2001} but also the construction of the network itself (\Cref{fig:empirical}.\textbf{f}).
Pair-wise connections among all co-authors of a paper form many triangles among them, leading to high CCF values.
Trends are increasing for the collaboration networks, but decreasing for citations.

\subsection*{Underlying behavioral mechanisms}

\begin{figure}[t]
    \centering
    \includegraphics[width=\textwidth]{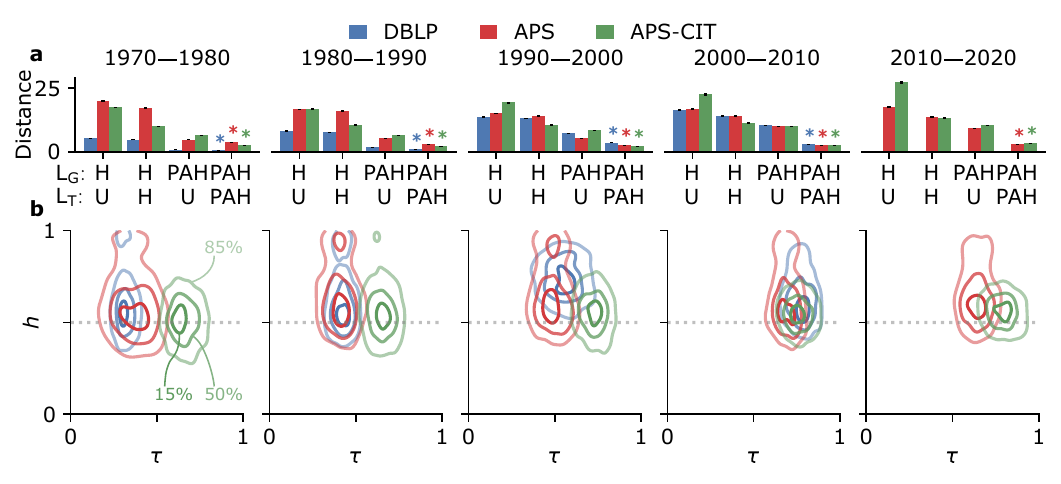}
    \caption{\label{fig:inference}\textbf{Likelihood-free inference of PATCH variant and parameters}. (\textbf{a}) Euclidean distance between observed and simulated network statistics for the empirical networks (color) and the PATCH model variants over decades.
    The best model variant is chosen based on the minimum distance (marked by a star; see~\nameref{sec:methods}).
    \model{PAH}{PAH} performs best in all cases, indicating that both the global and triadic closure target selection is biased by popularity and gender-homophily.
    (\textbf{b}) Approximate joint posterior distributions of the homophily parameter $h$ and triadic closure probability $\tau$ for \model{PAH}{PAH} over time with the contours incorporating $85\%$, $50\%$ and $15\%$ of the posterior samples.
    The distributions show a moderate gender-homophily and triadic closure preferences.
    }
\end{figure}

To estimate which behavioral biases could explain the observed inequalities in the empirical networks, we apply Approximate Bayesian Computation (ABC) to estimate the posterior parameter distributions of $h$ and $\tau$ that best explain the observed network statistics (see~\nameref{sec:methods} for details on the inference approach).~\cite{gutmann.etal_bayesianoptimizationlikelihoodfree_2016}
ABC uses the distance between summary statistics of the observed and simulated networks to sample from an approximate posterior distribution of the model parameters.

After validating the inference approach on simulated data (see~\Cref{fig:si_confusion,fig:si_post_hh,fig:si_post_hu,fig:si_post_pahpah,fig:si_post_pahu} and~\nameref{sec:methods} for details), we apply it to the empirical networks.
We choose the model variant of PATCH that minimizes the Euclidean distance between the observed and simulated network statistics (\Cref{fig:inference}.\textbf{a}).
There is strong variation in how well model variants fit the observed inequalities.
Over time, the second best model variant \model{PAH}{U} becomes less viable in all datasets, indicating that the triadic closure selection is increasingly probable to be biased by homophily and preferential attachment.
In all datasets and decades, we find that homophily and preferential attachment in the global and triadic closure links \model{PAH}{PAH} reproduces the observed network statistics and inequalities best.
In contrast to existing studies,~\cite{kossinets.watts_originshomophilyevolving_2009} popularity and similarity also seem to shape linking selection among local authors.
%Popularity and similarity biases in how collaborations and citations are created are the most likely underlying mechanisms of the observed gender disparities in the selected scientific networks under the PATCH model.

Samples of the approximate posterior distributions of parameters $h$ and $\tau$ for the \model{PAH}{PAH} models estimate the latent behavior that best explains the observed inequalities (\Cref{fig:inference}\textbf{b}, see also~\nameref{sec:methods} and~\Cref{fig:si_pred_ei,fig:si_pred_gini_m,fig:si_pred_gini_M,fig:si_pred_mw,fig:si_pred_ccf,fig:si_pred_gini_comp}).
Both collaborations and citations in most decades are mildly gender-homophilious and show moderate triadic closure tendencies which slightly increase over time for all networks.
This suggests that scientists, increasingly, base their decisions to collaborate or cite among friends of friends, choosing based on the popularity and gender of authors (implicit or explicitly).
Adding to the ongoing discussion on gender biases in citation practices,~\cite{dworkin.etal_extentdriversgender_2020,jaramillo.etal_systematiccomparisongender_2025} this suggests that biases are present both in the global and local selection of citations.

Our inference model reproduces the high levels of segregation, the within-group inequality ratio, and inequity in the empirical networks (\Cref{fig:si_pred_ei,fig:si_pred_gini_m,fig:si_pred_gini_M,fig:si_pred_mw,fig:si_pred_gini_comp}). Returning back to our simulation results for \model{PAH}{PAH} (\Cref{fig:segregation,fig:inequality,fig:inequity_inequality}), this suggests that reducing gender homophily could improve women's segregation and degree disadvantage.
At the same time, this would decrease the degree inequality among men, at the expense of an increase for women, suggesting that only a few, already popular women would benefit from the majority group's attention shift.
This trade-off stabilizes the global degree inequality which can only be reduced by decreasing the effect of preferential attachment or triadic closure.
To move the empirical networks towards a more equal state in all inequality dimensions, a simultaneous reduction of all mechanisms is required.

\starsection{Discussion}{sec:discussion}
As a model combining the common link formation mechanisms of preferential attachment, homophily, and triadic closure, PATCH produces a complex interplay of network inequalities.
We find that the presence of unbiased triadic closure can have a moderating effect on network segregation, but cannot confirm its segregating effect under homophily identified in previous studies.~\cite{abebe.etal_effecttriadicclosure_2022}
Unbiased triadic closure further reduces global and within-group degree inequality if links are otherwise chosen based on preferential attachment.
With the exception of extreme or no homophily, inequality is always higher in one of the two groups.
Inequity, on the other hand, shows that the minority is disfavored in homophilic settings but favored in heterophilic ones.
While preferential attachment increases inequity in all cases, triadic closure tends to reduce it.
Combining inequality and inequity, we find that the group accumulating more links is also the one with higher inequality.
Only a few nodes in the favored group receive a large share of the links, while the majority of the group is left with lower degree visibility.

% Applying the model to fifty years of empirical scientific collaboration and citation networks, we find that global preferential attachment and gender-homophily with moderate triadic closure can most closely reproduce the observed gender inequalities.
% This variant shows reduced inequality, inequity and segregation only when both homophily and triadic closure are reduced.

Applying model inference to fifty years of empirical scientific collaboration and citation networks, we find that the model variant in which both global and local (triadic closure) links are biased by preferential attachment and moderate gender homophily best reproduces the observed gender inequalities.
These findings are valuable in guiding network interventions.
Reducing inequality, inequity, and segregation in networks following these dynamics requires simultaneously lowering both homophily and triadic closure.
However, improvements to group-level inequalities can negatively impact individuals.
For instance, triadic closure is linked to individual performance in scientific production and citations received,~\cite{bachmann.etal_cumulativeadvantagebrokerage_2024} and reducing preferential attachment would naturally disfavor popular scientists.

% \subsection{Limitations}
% - comprehensive but not complete:
% 	- local homophily different from global homophily
% 	- varying homphily between groups
% 	- only generative model
By considering the presence and absence of a collection of link formation mechanisms, we disentangle potential causal drivers of each mechanism on the observed network structure.
However, our model is not exhaustive.
For example, we fix an identical homophily value for both groups.
In reality, one group might be more homophilic than another.
Empirical evidence suggests that homophilic tendencies depend on group size and can vary between groups.~\cite{currarini.etal_identifyingrolesracebased_2010}
Homophilic preference could further divert between global and triadic closure target selection, modeling differences in how people prefer linking to others based on the context in which they meet them.
This could be tied to empirical findings that suggest that homophilic tendencies decrease with network proximity.~\cite{kossinets.watts_originshomophilyevolving_2009}
% Most scientific networks studied here have a higher inequality among the minority group and inequity disfavoring them.
% Allowing for asymmetric homophily could enable the model to capture these patterns, further improving its capability to reproduce the interplay of empirical inequalities.
Tunable and asymmetric preferential attachment could bring the simulated degree inequality closer to that of empirical networks.
Lastly, our network model is only a simplification of reality.
Our model is a growth model, meaning that we only consider the formation of new links and do not alter established links.
In many empirical social networks, individuals may decide to break ties with others or exit the network altogether.
While improving the model's realism is valuable, adding more parameter variations and asymmetries would also increase its complexity and thus complicate interpretation.

% - criticize empirical data application
%   - combination of higher minority inequality and inequity cannot be captured by model
%   - high tau values also through network generation
% \subsection{Implications}
% - effect of triadic closure nuanced
% 	- strong inter-dependence on other mechanisms
% 	- go through some examples?
% 	- implies necessity to assess which mechanisms are at play before
% 		- suggesting interventions
% 		- claiming beneficial or harmful effect on minority
Mechanistic models like PATCH have the potential to explain the emergence of network structure,~\cite{barabasi.albert_emergencescalingrandom_1999} inequalities~\cite{karimi.etal_homophilyinfluencesranking_2018,espin-noboa.etal_inequalityinequitynetworkbased_2022} and the effects of interventions.~\cite{neuhauser.etal_improvingvisibilityminorities_2023}
If policy making aims to reduce network-based inequalities, it is crucial to understand the interplay of mechanisms that cause them.
Our results suggest that the effect of triadic closure is nuanced and depends on the presence of other mechanisms.
The implementation of interventions, such as recommendation algorithms boosting triadic closure or preferential attachment by suggesting friends of friends, or popular people, should therefore carefully consider the present context to balance the mechanisms and desired outcomes.
For example, in a homophilic environment and in the presence of preferential attachment, such as academic collaboration or citation, the introduction of a triadic closure based recommendation algorithm could slightly reduce inequity at the cost of increased inequality if the algorithm was biased by homophily.
In contrast, an algorithm agnostic to both popularity and homophily could potentially reduce both inequality and inequity.

\starsection{Methods}{sec:methods}

\subsection*{Model specification}
As an extension to the Barab\'asi-Albert model, PATCH is initialized with $m$ fully connected nodes.~\cite{barabasi.albert_emergencescalingrandom_1999}
Each of the remaining $N-m$ nodes joins the network one after the other and links to $m$ previously added nodes.
Group assignments are drawn randomly with probability $\fmin$ for each node.
For each link, the node can choose globally among all previously added nodes with probability $1 - \tau$, following prior triadic closure model variants.
Otherwise, with triadic closure probability $\tau$, we limit available targets nodes to neighbors of neighbors.~\cite{asikainen.etal_cumulativeeffectstriadic_2020,bianconi.etal_triadicclosurebasic_2014,holme.kim_growingscalefreenetworks_2002}
Note that, because a node initially has no connections, its first link is always global.
Among the chosen target set of nodes, the selection is then based on a random (uniform) choice (\compound{U}), homophily (\compound{H}), or preferential attachment and homophily (\compound{PAH}).
The probability of forming a link between nodes $i$ and available target nodes $j$ is then given by
\begin{equation}
    \Pi_{ij} = \frac{p_{ij}}{\sum_{n < i} p_{in}}
\end{equation} where $p_{ij}$ depends on the chosen link formation mechanisms.

In the uniform case (\compound{U}), we fix $p_{ij} = 1$.
For preferential attachment and homophily (\compound{PAH}), we follow a variant of the homophily model~\cite{karimi.etal_homophilyinfluencesranking_2018} and set $p_{ij} = h_{ij}k_j$, where $k_j$ is the degree of node $j$ and $h_{ij}$ depends on the homophily parameter $h \in (0,1)$ and the minority/majority attribute of nodes $i$ and $j$.
If $i$ and $j$ belong to the same group, $h_{ij} = h_{ji} = h$; otherwise, $h_{ij} = 1 - h$.
Values of $h > 0.5$ result in homophilic tendencies, preferring in-group links, while $h < 0.5$ leads to heterophily.
Note that $h=0.5$ neutralizes the effect of homophily, corresponding to preferential attachment without homophily.
This recovers the triadic closure only variants~\cite{holme.kim_growingscalefreenetworks_2002,bianconi.etal_triadicclosurebasic_2014} and the original Barab\'asi-Albert model~\cite{barabasi.albert_emergencescalingrandom_1999} if triadic closure is also disabled ($\tau = 0$).
For homophily without preferential attachment (\compound{H}), we neutralize the degree effect by setting $p_{ij} = h_{ij}$.
If not stated otherwise, we fix $N = 5\,000$, $\fmin = 0.2$, $m = 3$, simulate each model $100$ times, and vary $h$ and $\tau$ as control parameters.

Our model definition is flexible to varying link formation mechanisms $\mathrm{L_{G,T}} \in \{\compound{U}, \compound{H}, \compound{PAH}\}$ when forming global $\mathrm{L_G}$ or triadic closure edges $\mathrm{L_T}$.
For instance, nodes may be driven by homophily and preferential attachment when connecting globally to any existing node of the network ($\mathrm{L_G} = \compound{PAH}$), but then decide to connect to their local neighborhood without any group or popularity preference ($\mathrm{L_G} = \compound{U}$).
To reduce the number of model combinations while still retaining the flexibility of biasing triadic closure, we restrict our analysis to models in which triadic closure links follow either the unbiased case ($\mathrm{L_T} = \compound{U}$) or whatever mechanism is specified for global links ($\mathrm{L_T} = \mathrm{L_G}$), including the respective homophily parameterization $h$.

\subsection*{Empirical networks}
To account for representational and behavioral changes in time, such as the increase of the fraction of women~\cite{zappala.etal_genderdisparitiesdissemination_2024} (see also \Cref{fig:empirical}.\textbf{a}) or decreasing gender homophily in collaboration,~\cite{sajjadi.etal_unveilinghomophilypool_2024} we consider decade-long network snapshots, starting between 1970 and 2010 for APS and APS-CIT, and between 1970 and 2000 for DBLP.
In PATCH, all existing nodes are available targets for linking.
Although this is realistic when citing older papers, authors may no longer be available for collaboration after retiring.
In the collaboration networks (APS and DBLP), we thus only consider active authors as those who continue to publish after the end of the respective decade.
The post-decade observation period for observing these publications covers one year for APS and eight years for DBLP and it mainly affects only the last decade of each dataset.
Each snapshot consists of the scientists for whom a gender label could be inferred from their names using an open source inference approach at a 95\%-certainty threshold.~\cite{buskirk.etal_opensourceculturalconsensus_2023}
The disambiguation of scientists, tracking their publications over time, is provided by rule-based solutions for the APS~\cite{sinatra.etal_quantifyingevolutionindividual_2016,bachmann.etal_cumulativeadvantagebrokerage_2024} and DBLP datasets.~\cite{dblpteam_dblpcomputerscience_2016}
In the collaboration networks (APS and DBLP), the remaining authors are then linked pairwise if they co-authored a paper during the respective decade.
In the citation network APS-CIT nodes are papers, linked if one paper cites another.
We leave the inclusion of directionality of PATCH for future work and ignore it for this application.
In Physics, the order of author names on a paper typically signals the author's contribution.
Aligning with previous research, we are interested in gendered citation inequalities with regards to the first authors and thus label a paper to belong to the minority group if its first author was labelled as a female scientist.

\subsection*{Likelihood-free inference}
To learn the underlying link formation mechanisms that best explain the inequalities of the empirical networks, we want to estimate the posterior distribution of the model parameters $h$ and $\tau$
\begin{equation}
    p(h, \tau | G) = \frac{p(G | h, \tau) p(h, \tau)}{p(G)},
\end{equation}
where $p(h, \tau)$ is the prior distribution of the model parameters and $p(G | h, \tau)$ is the likelihood of the observed networks $G$ given the model parameters.
However, the likelihood term is intractable for our generative model.
Approximate Bayesian computation (ABC) is a likelihood-free inference method that allows us to estimate the posterior distribution of the model parameters by comparing simulated and observed networks by a selection of summary statistics.~\cite{rubin_bayesianlyjustifiablerelevant_1984,lintusaari.etal_fundamentalsrecentdevelopments_2017,marin.etal_approximatebayesiancomputational_2012}
Beyond the metrics of segregation (EI), inequity (Mann-Whitney test statistic) and within-group inequalities ($\mathrm{Gini_{min}}$, and $\mathrm{Gini_{maj}}$), we inform the inference with the average local clustering coefficient (CCF).~\cite{newman_networksintroduction_2010}
By measuring the fraction of closed triangles, we expect the CCF to be informative about the triadic closure parameter $\tau$ which we showed to have a moderate effect on the other metrics.
We use the local clustering coefficient to decrease the influence of the network size on the summary statistics, which we fix to $N_{\text{sim}} = 500$ nodes to balance computational costs.
The global degree inequality ($\mathrm{Gini}$) is not used for the inference to reduce the focus on degree-based inequalities as they are already captured by the two within-group measures.
The number of links per node $m$ matches the average degree of the empirical networks with a lower bound of $m \geq 2$ (see~\Cref{ssec:inference_m}).
The prior distribution is uniform for both parameters $h,\tau \sim \mathrm{Uniform}(0, 1)$ and the minority fraction $\fmin$ is chosen directly from the empirical networks (see~\Cref{fig:empirical}.\textbf{a}).
Using the \texttt{elfi} Python package,~\cite{lintusaari.etal_elfienginelikelihoodfree_2018} we apply seven rounds of Sequential Monte Carlo sampling with adaptive distance weighting~\cite{prangle_adaptingabcdistance_2017} to draw 1,000 posterior samples of the model parameters $h$ and $\tau$.

We evaluate this setup using a three-step process.
The first two steps apply a synthetic evaluation.
We simulate PATCH networks with varying model variants and known parameters $h$ and $\tau$, averaging the summary statistics over 100 simulations.
In the first evaluation step, we aim to retrieve the model variant used during the simulations.
Each fit yields a distribution of distances between the observed summary statistics and the statistics corresponding to the accepted posterior sample.
We select the model variant with the lowest distance distribution compared to the unified distance distribution of all other variants.
Our approach correctly classifies most model variants, but confuses local link formation mechanisms in the absence of triadic closure, or under extreme or neutral homophily values (see \Cref{fig:si_confusion} and its caption for a discussion).

The second step evaluates whether the inferred posterior distributions of the selected model variant center around the true parameters.
We identify good agreement for all simulated variants, with increasing uncertainty for more expressive models which include \compound{PAH} in its link formation mechanisms (\Cref{fig:si_post_hh,fig:si_post_hu,fig:si_post_pahu,fig:si_post_pahpah}).

As a third evaluation step, we perform predictive checks to test whether the selected inferred model can reproduce the empirical network inequalities.
We compute summary statistics averaged across 100 PATCH network simulations for each of the 1,000 posterior sample pairs $(h, \tau)$ and compare them to the empirical statistics.
Although PATCH can reproduce empirical segregation, inequity, and clustering surprisingly well (\Cref{fig:si_pred_ei,fig:si_pred_mw,fig:si_pred_ccf}), the group-wise degree inequalities are not well captured (\Cref{fig:si_pred_gini_m,fig:si_pred_gini_M}).
Because the strength of preferential attachment is fixed in PATCH, it cannot match the strong degree inequality in the empirical networks.
However, the model reproduces the relative ratio of degree inequality between the two groups, even if the measure was not used for the inference (\Cref{fig:si_pred_gini_comp}).
Given PATCH's simplicity, we consider this a reasonable result which also informs us about model limitations and the need for further model extensions.

\starsection{Data Availability}{sec:data_avail}
The APS and APS-CIT datasets~\cite{americanphysicalsociety_americanphysicalsociety_2023} are available by request from \url{https://journals.aps.org/datasets/aps-cit}.
The DBLP dataset~\cite{dblpteam_dblpcomputerscience_2016,ley_dblpcomputerscience_2002} is freely available at \url{https://dblp.org/}.
For this analysis, a pre-processed network version of DBLP is used, which is available through the \texttt{konect} network repository~\cite{kunegis_konectkoblenznetwork_2013} at \url{http://konect.cc/networks/dblp_coauthor/}.

\starsection{Code Availability}{sec:code_avail}
The code to reproduce the results of this paper is available at \url{https://github.com/mannbach/patch}.
The repository is linked to a Zenodo archive at \url{https://doi.org/10.5281/zenodo.17160884}, which contains the simulated networks and aggregated statistics.
The PATCH model is further implemented in version \texttt{2.0.0a2} of the \texttt{netin} Python package, which is available at \url{https://pypi.org/project/netin/2.0.0a2/}.

\bibliography{Patch}

\starsection{Acknowledgements}{sec:acknowledgements}
The authors thank Eric Dignum for helpful discussions on the likelihood-free inference approach, Liuhuaying Yang for her support on the presentation of results, and Tobias Schumacher for his feedback on the model notation.

\textbf{Funding}
J.B. was supported by the Austrian Science Promotion Agency FFG under project No. 873927 ESSENCSE.
J.B. is a recipient of a DOC Fellowship of the Austrian Academy of Sciences at the Complexity Science Hub.
A poster presentation of this work was supported by the \"OFG ``Internationale Kommunikation" travel grant.
S.M.G. and F.K. were funded by the European Union under the Horizon Europe MAMMOth project, Grant Agreement ID: 101070285, and by the Austrian Research Promotion Agency (FFG) under project No. 873927.
L.E.N. was supported by the Vienna Science and Technology Fund WWTF
under project No. ICT20-079.

\starsection{Author contributions}{sec:author_contributions}
% Based on CRediT https://www.elsevier.com/researcher/author/policies-and-guidelines/credit-author-statement
% F.K. and J.B. conceived the study.
All authors conceived the study and contributed to the design of the model.
J.B. implemented the model code, performed the simulations, designed, executed the inference and its validation, collected and processed the empirical data, wrote the original draft and visualized the results.
% S.M.G. presented the model and results at the Deutsche Physiker Gesellschaft (DPG) Spring Meeting 2024.
% J.B. presented a poster at the 2025 International Conference on Computational Social Science (IC$^2$S$^2$) in Norrk\"oping, Sweden.
F.K. supervised the project.
All authors contributed to the interpretation and discussion of the results, and the writing of the manuscript.

\starsection{Competing interests}{sec:competing_interests}
The authors declare no competing interests.

\newpage

\starsection{Supplementary information}{sec:supplementary_material}

% Reset subsection counter and add S prefix
\setcounter{subsection}{0}
\renewcommand{\thesubsection}{S\arabic{subsection}}

\subsection{Inferring $m$}
\label{ssec:inference_m}
We infer the number of links per node $m$ used in the inference simulations by matching the average degree between the observed empirical and the simulated networks $\langle k_\semp \rangle \approx \langle k_\ssim \rangle$.
The average simulation degree
\begin{equation}
    \langle k_\ssim \rangle = \frac{2L_\ssim}{N_\ssim}
\end{equation}
depends on the total number of nodes $N_\ssim$ and links $L_\ssim$.
To balance computational costs, we fix the number of nodes to $N_\ssim = 500$.
The number of links $L_\ssim$ in PATCH depends on the number of links per node $m$ and the number of nodes $N_\ssim$ as
\begin{align}
    L_\ssim &= \underbrace{\frac{m(m-1)}{2}}_{\text{initial clique}} + \underbrace{(N_\ssim - m) m}_{\text{added nodes}}\\
            &= m N_\ssim - \frac{m(m+1)}{2}.
\end{align}
We infer $m$ by equalizing the average degree of the empirical networks $\langle k_\semp \rangle$ with the average degree of the simulated networks $\langle k_\ssim \rangle$
\begin{align}
    \langle k_\semp \rangle &= \langle k_\ssim \rangle\\
                            &= \frac{2}{N_\ssim}L_\ssim.
\end{align}
Substituting the expression for $L_\ssim$ and rearranging the terms yields
\begin{align}
    \frac{2}{N_\ssim} (m N_\ssim - \frac{m(m+1)}{2}) - \langle k_\semp \rangle &= 0\\
    2m - \frac{m(m+1)}{N_\ssim} - \langle k_\semp \rangle &= 0
\end{align}
Multiplying by $N_\ssim$ and additional rearranging gives the quadratic equation
\begin{equation}
    m^2 - m (2N_\ssim - 1) + \langle k_\semp \rangle N_\ssim = 0
\end{equation}
which we solve for $m$ using the quadratic formula
\begin{equation}
    m_{+,-} =\frac{1}{2} \left( 2N_\ssim - 1 \pm \sqrt{(2N_\ssim - 1)^2 - 4 \langle k_\semp \rangle N_\ssim} \right).
\end{equation}
Because $m \leq N_\ssim$ is a strict upper bound for $m$, we only consider the subtraction solution.
As triadic closure links are only formed between neighbors of neighbors, we finally require $m \geq 2$
\begin{equation}
\label{eq:initial_links}
    m = \max\left(2,\ \frac{1}{2} \left[ 2N_\ssim - 1 - \sqrt{(2N_\ssim - 1)^2 - 4  \langle k_\semp \rangle N_\ssim}\right]\right),
\end{equation}
and round the solution to the nearest integer.

% Note that the term with the square root can be rewritten as $2 N_{\ssim}\sqrt{1-\frac{\langle k_\semp \rangle}{N_{\ssim}}}$ by taking into account that $N_{\ssim}>>1$. Since also $N_{\ssim}>> \langle k_\semp \rangle$, we can further approximate the square root by $\sqrt{1-\frac{\langle k_\semp \rangle}{N_{\ssim}}} =1 - \frac{1}{2}\frac{\langle k_\semp \rangle}{N_{\ssim}} + O(\left[ \frac{\langle k_\semp \rangle}{N_{\ssim}}\right]^2)$. If we plug this expression into Eq. \eqref{eq:initial_links}, we would obtain that $m \approx \frac{\langle k_\semp \rangle - 1}{2}$ or, reciprocally, that $\langle k_\semp \rangle = 2m+1$.

% \newpage
% \FloatBarrier
% \starsection{Supplementary tables}{sec:supplementary_tables}
% \setcounter{table}{0}
% \renewcommand{\thetable}{S\arabic{table}}

% Start new label counts
\newpage
\FloatBarrier
\setcounter{figure}{0}
\renewcommand{\thefigure}{S\arabic{figure}}

\starsection{Supplementary figures}{sec:supplementary_figures}
\begin{figure}[ht]
    \centering
    \includegraphics[width=\textwidth]{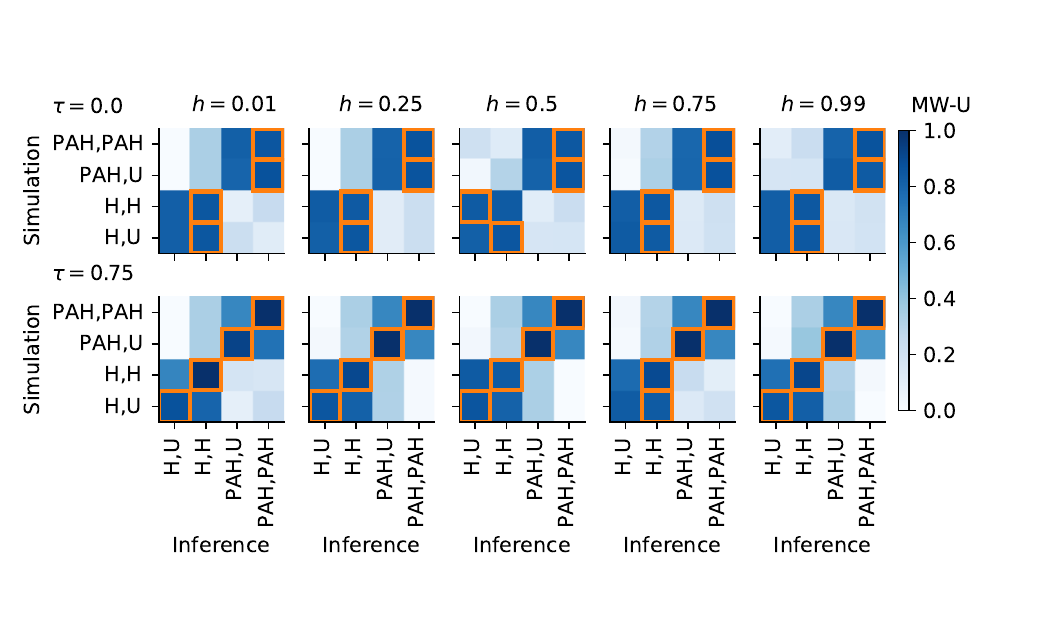}
    \caption{\label{fig:si_confusion}\textbf{Confusion matrix for model selection}.
    We simulate networks based on varying homophily $h$ (columns), triadic closure $\tau$ (rows) parameters, and link formation mechanisms $\mathrm{L_G}$ and $\mathrm{L_T}$ (matrix rows).
    We then fit the model to the simulated data and plot the Mann-Whitney test statistic (MW-U) comparing the sampled distances to the unified distribution of all other model variants' distance samples.
    Higher values indicate smaller distances compared to the other models, that is, a better fit.
    We then select the model with the highest MW-U value for each true model (column-based selection, marked by orange squares).
    A perfect fit would be indicated by a diagonal line of orange squares.
    For $\tau=0$ (upper row), the selection confuses variants with identical global selection mechanisms $\mathrm{L_G}$ but different triadic closure mechanisms $\mathrm{L_T}$ which is expected due to the lack of triadic closure.
    For $\tau=0.75$ (lower row), the selection is more accurate. It only confuses the true model \model{H}{U} with \model{H}{H}, under moderate homophily $h=0.75$.}
\end{figure}

\begin{figure}
    \centering
    \includegraphics[width=\textwidth]{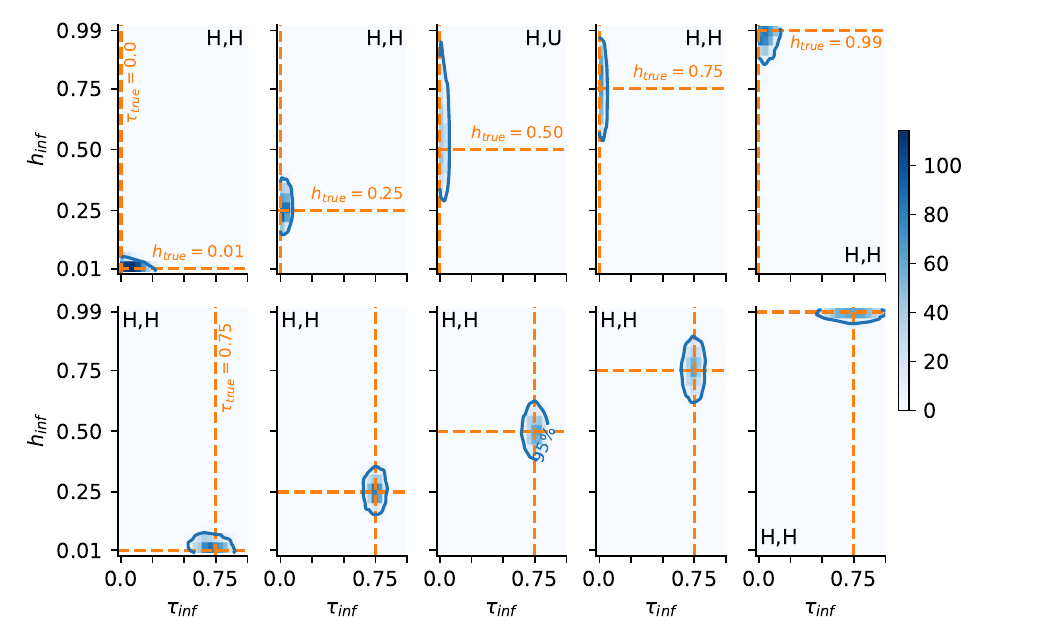}
    \caption{\label{fig:si_post_hh}\textbf{\model{H}{H} posterior distributions inference}.
    We simulate \model{H}{H} networks based on varying homophily $h_{true}$ and triadic closure $\tau_{true}$ parameters (orange lines) and infer the posterior distributions of the homophily $h_{inf}$ and triadic closure $\tau_{inf}$ parameters (blue heatmap) using the best fitting model variant (model labels in corners).
    95\% of the probability mass is contained in the area marked by the blue contours.
    Ideally, the posterior distribution should be centered closely around the intersecting true parameters (orange lines).
    }
\end{figure}

\begin{figure}
    \centering
    \includegraphics[width=\textwidth]{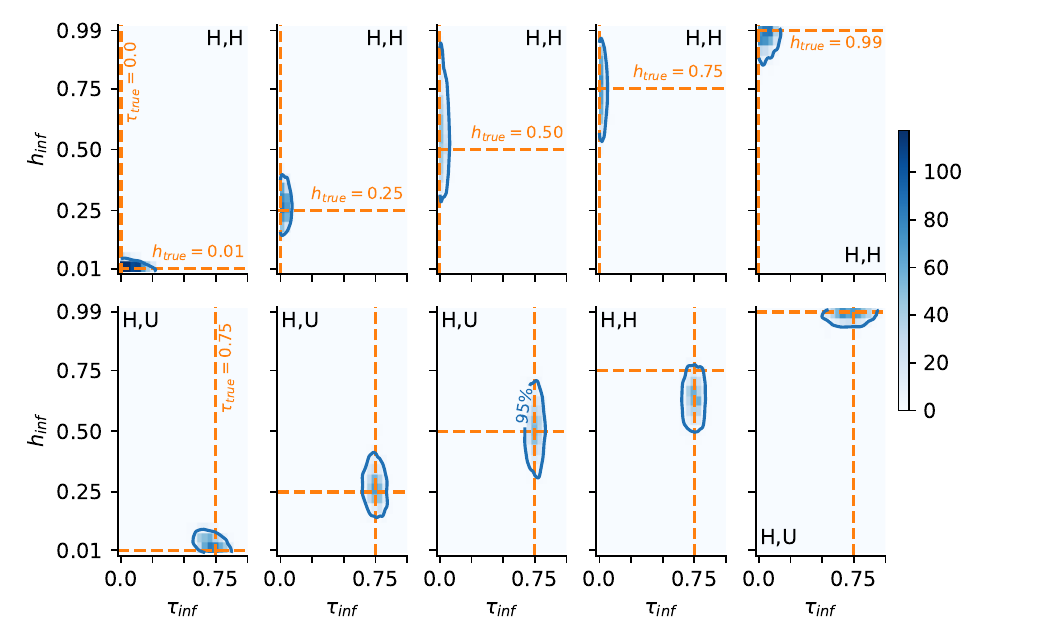}
    \caption{\label{fig:si_post_hu}\textbf{\model{H}{U} posterior distributions inference}.
    We simulate \model{H}{U} networks based on varying homophily $h_{true}$ and triadic closure $\tau_{true}$ parameters (orange lines) and infer the posterior distributions of the homophily $h_{inf}$ and triadic closure $\tau_{inf}$ parameters (blue heatmap) using the best fitting model variant (model labels in corners).
    95\% of the probability mass is contained in the area marked by the blue contours.
    Ideally, the posterior distribution should be centered closely around the intersecting true parameters (orange lines).
    }
\end{figure}

\begin{figure}
    \centering
    \includegraphics[width=\textwidth]{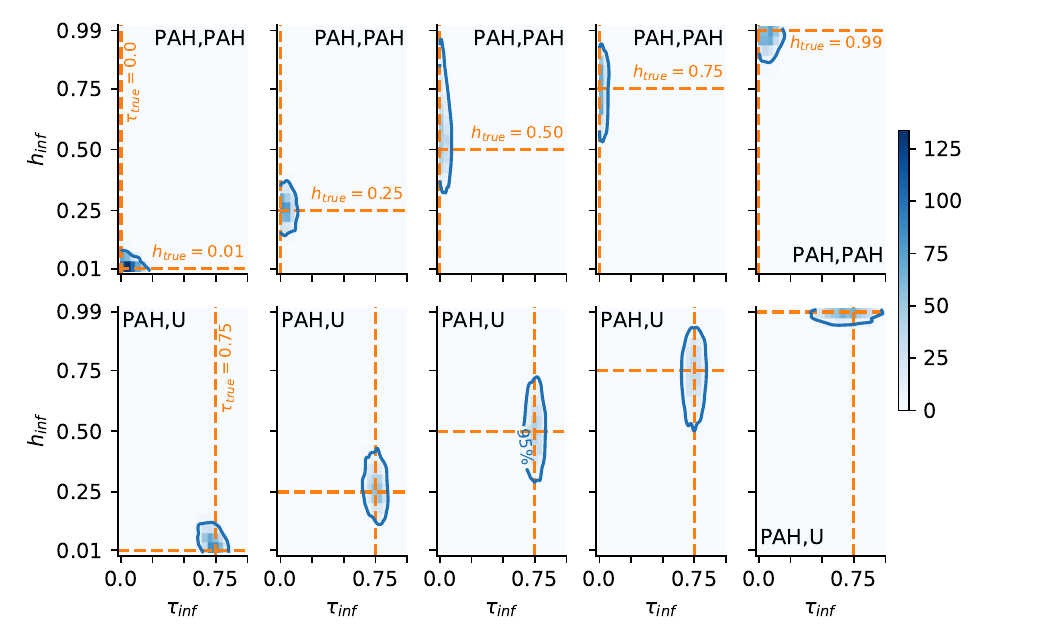}
    \caption{\label{fig:si_post_pahu}\textbf{\model{PAH}{U} posterior distributions inference}.
    We simulate \model{PAH}{U} networks based on varying homophily $h_{true}$ and triadic closure $\tau_{true}$ parameters (orange lines) and infer the posterior distributions of the homophily $h_{inf}$ and triadic closure $\tau_{inf}$ parameters (blue heatmap) using the best fitting model variant (model labels in corners).
    95\% of the probability mass is contained in the area marked by the blue contours.
    Ideally, the posterior distribution should be centered closely around the intersecting true parameters (orange lines).
    }
\end{figure}

\begin{figure}
    \centering
    \includegraphics[width=\textwidth]{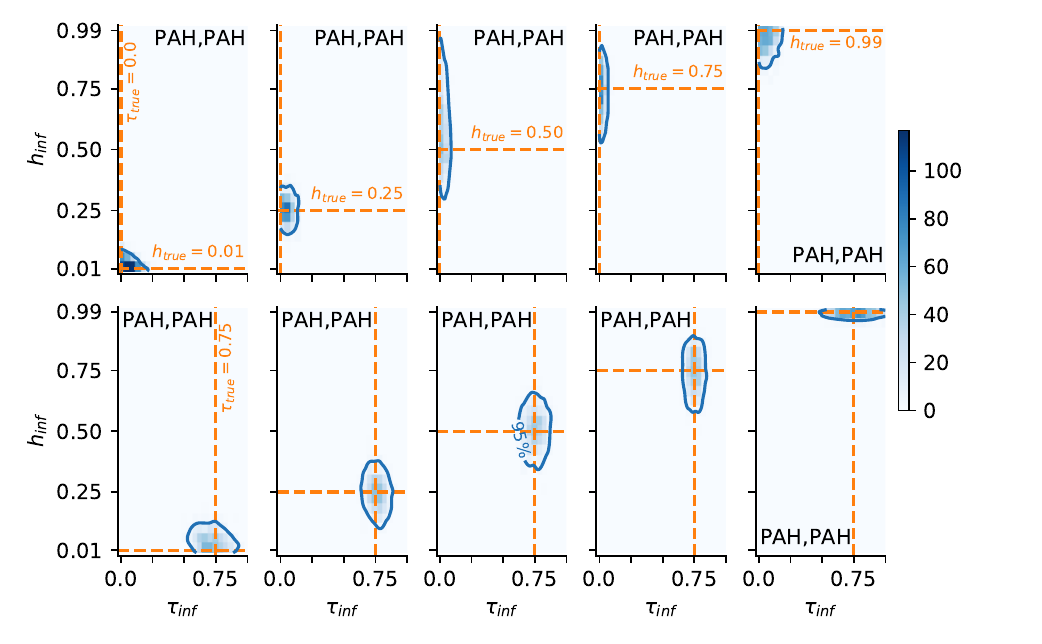}
    \caption{\label{fig:si_post_pahpah}\textbf{\model{PAH}{PAH} posterior distributions inference}.
    We simulate \model{PAH}{U} networks based on varying homophily $h_{true}$ and triadic closure $\tau_{true}$ parameters (orange lines) and infer the posterior distributions of the homophily $h_{inf}$ and triadic closure $\tau_{inf}$ parameters (blue heatmap) using the best fitting model variant (model labels in corners).
    95\% of the probability mass is contained in the area marked by the blue contours.
    Ideally, the posterior distribution should be centered closely around the intersecting true parameters (orange lines).
    }
\end{figure}

\begin{figure}
    \centering
    \includegraphics[width=\textwidth]{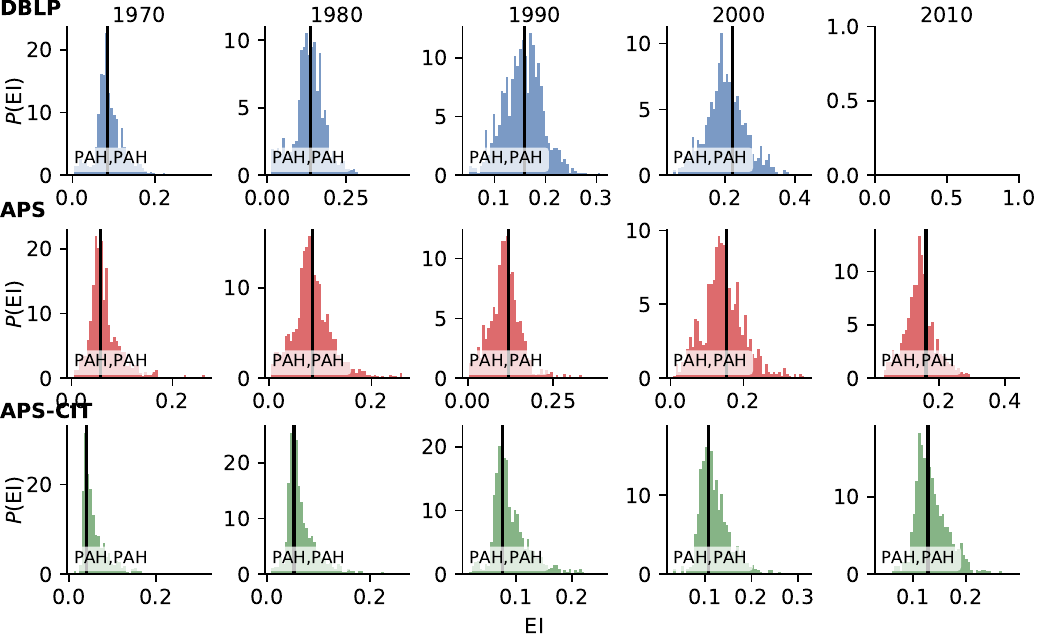}
    \caption{\label{fig:si_pred_ei}\textbf{$\mathrm{EI}$-index predictive analysis}.
    For all $h$ and $\tau$ approximate posterior sample pairs, we compute the $\mathrm{EI}$-index average of 100 PATCH simulations.
    We compare the histogram over all 1,000 pairs to the observed $\mathrm{EI}$-index (vertical line) to see if PATCH can reproduce it.
    Network segregation is well captured by PATCH in all datasets and decades.
    }
\end{figure}

\begin{figure}
    \centering
    \includegraphics[width=\textwidth]{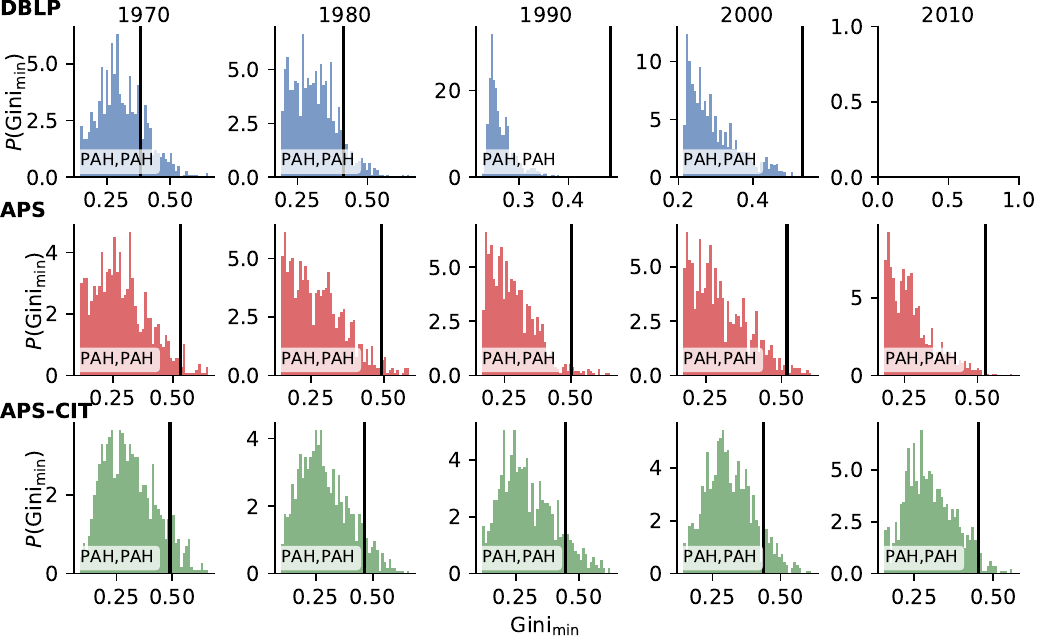}
    \caption{\label{fig:si_pred_gini_M}\textbf{$\mathrm{Gini_{min}}$ predictive analysis}.
    For all $h$ and $\tau$ approximate posterior sample pairs, we compute the $\mathrm{Gini_{min}}$ average of 100 PATCH simulations.
    We compare the histogram over all 1,000 pairs to the observed $\mathrm{Gini_{min}}$ (vertical line) to see if PATCH can reproduce it.
    PATCH cannot reproduce the degree inequality among minority notes observed in all networks.}
\end{figure}

\begin{figure}
    \centering
    \includegraphics[width=\textwidth]{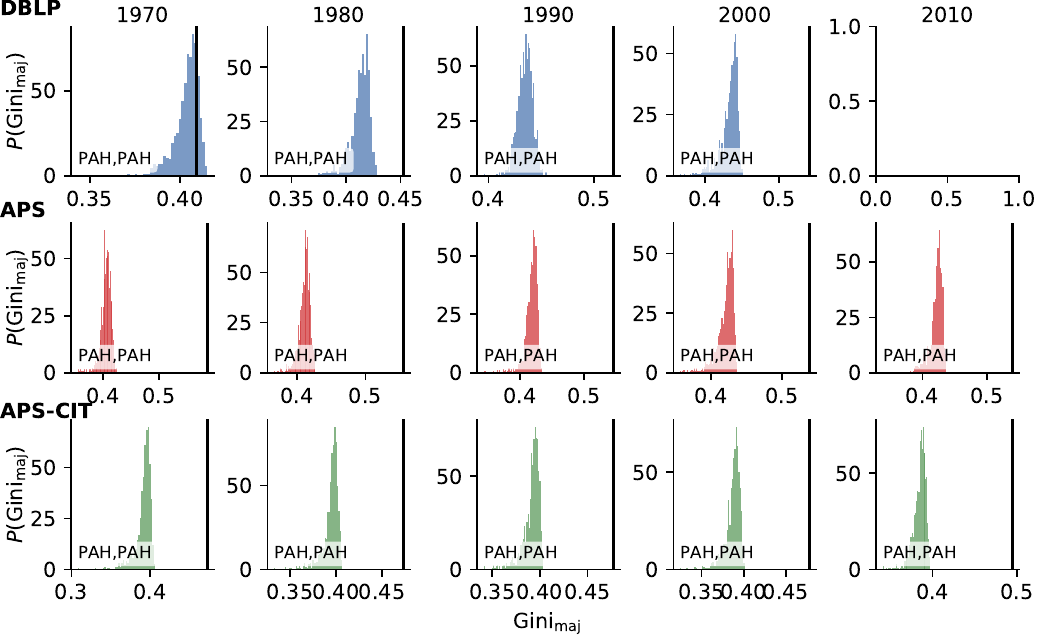}
    \caption{\label{fig:si_pred_gini_m}\textbf{$\mathrm{Gini_{maj}}$ predictive analysis}.
    For all $h$ and $\tau$ approximate posterior sample pairs, we compute the $\mathrm{Gini_{maj}}$ average of 100 PATCH simulations.
    We compare the histogram over all 1,000 pairs to the observed $\mathrm{Gini_{maj}}$ (vertical line) to see if PATCH can reproduce it.
    PATCH underestimates the degree inequality among majority notes consistently.}
\end{figure}
\begin{figure}
    \centering
    \includegraphics[width=\textwidth]{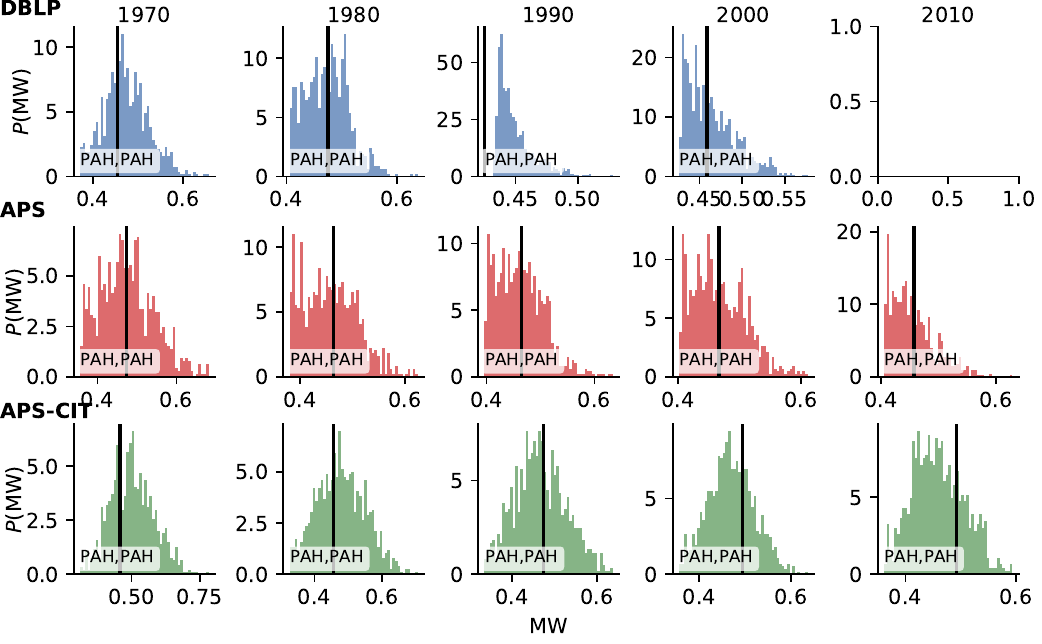}
    \caption{\label{fig:si_pred_mw}\textbf{$\mathrm{MW}$ predictive analysis}.
    For all $h$ and $\tau$ approximate posterior sample pairs, we compute the Mann-Whitney ($\mathrm{MW}$) test statistic average of 100 PATCH simulations.
    We compare the histogram over all 1,000 pairs to the observed $\mathrm{MW}$ (vertical line) to see if PATCH can reproduce it.
    PATCH reproduces the inequity among minority notes observed in most networks.}
\end{figure}

\begin{figure}
    \centering
    \includegraphics[width=\textwidth]{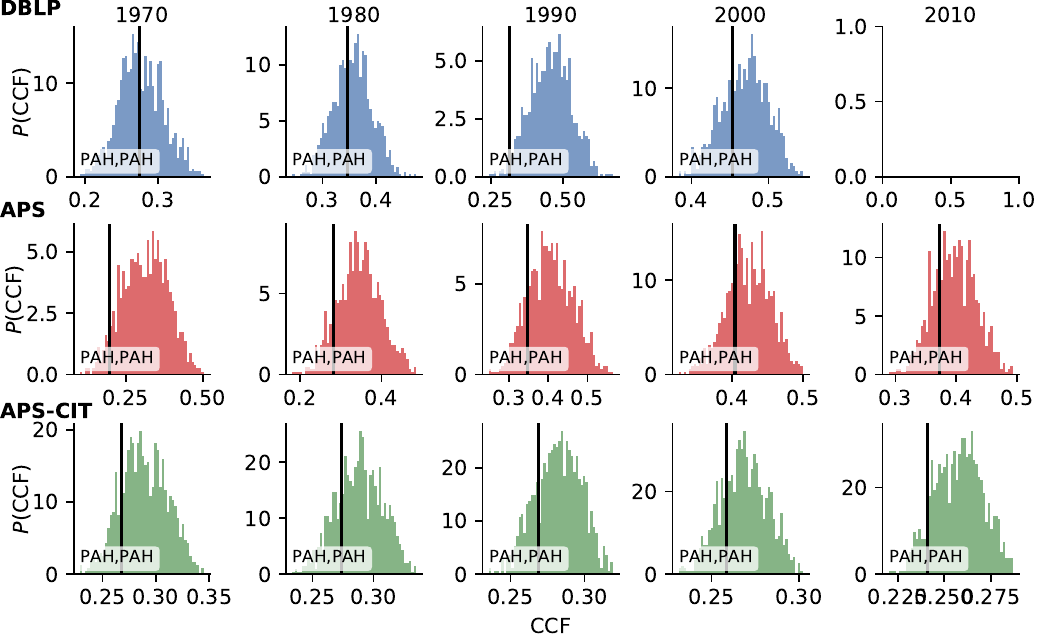}
    \caption{\label{fig:si_pred_ccf}\textbf{$\mathrm{CCF}$ predictive analysis}.
    For all $h$ and $\tau$ approximate posterior sample pairs, we compute the clustering coefficient ($\mathrm{CCF}$) average of 100 PATCH simulations.
    We compare the histogram over all 1,000 pairs to the observed $\mathrm{CCF}$ (vertical line) to see if PATCH can reproduce it.
    PATCH reproduces the observed network clustering in most cases.
    }
\end{figure}

\begin{figure}
    \centering
    \includegraphics[width=\textwidth]{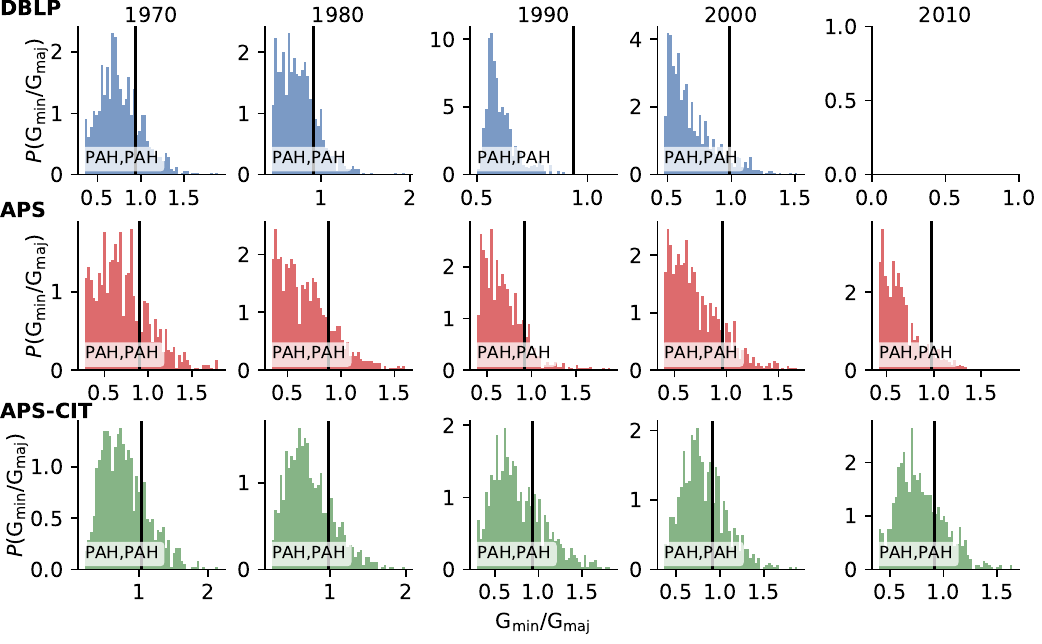}
    \caption{\label{fig:si_pred_gini_comp}\textbf{$\mathrm{Gini_{min} / Gini_{maj}}$ predictive analysis}.
    For all $h$ and $\tau$ approximate posterior sample pairs, we compute the $\mathrm{Gini_{min} / Gini_{maj}}$ average of 100 PATCH simulations.
    We compare the histogram over all 1,000 pairs to the observed $\mathrm{Gini_{min} / Gini_{maj}}$ (vertical line) to see if PATCH can reproduce it.
    PATCH roughly matches the within-group inequality ratio although it is not used during the inference.
    However, it consistently assigns a stronger inequality to the majority group.
    }
\end{figure}

\end{document}